\documentclass[seceq]{ptptex}

\usepackage{epsf}
\usepackage{graphicx}
\makeatletter

\newcommand{\beq}{\begin{equation}}
\newcommand{\eeq}{\end{equation}}
\newcommand{\beqn}{\begin{eqnarray}}
\newcommand{\eeqn}{\end{eqnarray}}

\newcommand{\pa}{\partial}

\newcommand{\varep}{\varepsilon}

\def\agt{\mathrel{\raise.3ex\hbox{$>$}\mkern-14mu\lower0.6ex\hbox{$\sim$}}}
\def\alt{\mathrel{\raise.3ex\hbox{$<$}\mkern-14mu\lower0.6ex\hbox{$\sim$}}}
\def\hbar{\mathrel{\raise.7ex\hbox{--}\mkern-14mu\hbox{$~h$}}}

\title{%        %You can use \\ for explicit line-break
Truncated Moment Formalism for Radiation Hydrodynamics
in Numerical Relativity%
}

\author{%       %Use \scshape  for the family name
Masaru \textsc{Shibata}$^1$, 
Kenta \textsc{Kiuchi}$^1$,
Yu-ichiro \textsc{Sekiguchi}$^{1,2}$,
Yudai \textsc{Suwa}$^1$
}

\inst{%     %Affiliation, neglected when [addenda] or [errata]
$^1$Yukawa Institute for Theoretical Physics, Kyoto University, Kyoto
606-8502, Japan \\ 
$^2$ Division of Theoretical Astronomy, National
Astronomical Observatory of Japan, 2-21-1, Osawa, Mitaka, Tokyo,
181-8588, Japan }

\abst{A truncated moment formalism for general relativistic radiation
hydrodynamics, based on the Thorne's moment formalism, is derived. The
fluid rest frame is chosen to be the fiducial frame for defining the
radiation moments. Then, zeroth-, first-, and second-rank radiation
moments are defined from the distribution function with a physically
reasonable assumption for it in the optically thin and thick limits.
The source terms are written, focusing specifically on the neutrino
transfer and neglecting higher harmonic angular dependence of the
reaction angle. Finally, basic equations for a truncated moment
formalism for general relativistic radiation hydrodynamics in a closed
covariant form are derived assuming a closure relation among the
radiation stress tensor, energy density, and energy flux, and a
variable Eddington factor, which works well.}

\begin{document}
\maketitle

\section{Introduction}

Radiation fields and their interaction with matter often play a
crucial role in many astrophysical contexts.  For example, the
critical roles of photon pressure during proto-star and massive-star 
formation and of neutrino heating and cooling in supernova core
collapse and explosion are well-known among many other phenomena.  To 
theoretically clarify these phenomena, it is necessary to solve
hydrodynamic equations as well as radiation transfer equations.  For
strictly handling the radiation transfer, it is necessary to
numerically solve the Boltzmann equation, taking into account the
absorption, emission, and scattering terms.  However, this equation
has 3+3+1 dimensional form (3 dimensions in real and phase spaces,
respectively, and 1 dimension in time), and furthermore, the time
scale for the interaction between matter and radiation is often
shorter than the dynamical time scale of the system. Thus, in the
current computational resources, it is not feasible to perform a
well-resolved numerical simulation with a sufficient grid resolution.
A certain approximate method incorporating key features of radiation
effects is often required in numerical astrophysics. In particular, no
useful formalism for multi-dimensional simulation in general
relativity has been well developed (but see Refs.~\citen{AS,Kip,CM}).
Note that in spherical symmetry, this equation is simplified to a
1+2+1 dimensional form as formulated in Ref.~\citen{MM}. Simulations
with similar formalism were performed in Ref.~\citen{GH}, and
subsequently, sophisticated simulations including the state-of-the-art
microphysics were achieved, e.g., in Refs.~\citen{Lieben,
Sumiyoshi}. However, the effort has been paid only to the spherical
symmetric simulations.  In this paper, we derive an approximate
formalism of radiation hydrodynamics in general relativity, in which a
numerical simulation will be feasible capturing the physically
important ingredients.

Historically, a popular method for approximate radiation hydrodynamics
is a flux-limited diffusion (FLD) method \cite{Miha}.  In this method,
the radiation flux density is in general assumed to be described by
the radiation energy density, and resulting evolution equation for the
radiation energy density becomes a diffusion-type equation in the
optically thick region (cf. \S~5).  In this case, the propagation
speed of characteristics may be larger than the speed of light,
although in general relativity, the causality must not be violated.
Another drawback of the FLD scheme is associated with the presence of
constraint equations (Hamiltonian and momentum constraints ) in the
initial value problem of general relativity: In numerical relativity
for multi-dimensional problems, we usually solve the evolution
equations of Einstein's equation and matter equations
self-consistently. As a result, the constraints are satisfied within a
numerical error. However, in the case that we do not solve the energy
and momentum equations for the radiation field self-consistently, the
constraints are violated. In the FLD method, one solves an equation
only for the radiation energy density component, and hence, the
constraints will be violated in general.

Truncated moment formalisms have been also proposed for an approximate
solution of radiation hydrodynamics \cite{AS,Kip}.  In this approach,
one derives a set of covariant equations for multi-pole moments
defined from the distribution function of radiation. Then, assuming
that higher-order moments may be neglected and imposing closure
relations, a closed covariant form of basic equations is derived. With
an appropriate choice of the closure relation, the causal relation can
be preserved, and furthermore, a solution of the radiation transfer in
the optically thick and thin limits can be derived from the resulting
equations.  In this paper, we derive a truncated moment formalism in
general relativity following the covariant formalism developed by
Thorne~\cite{Kip}. In addition, we derive a closed
coordinate-independent formalism including the absorption, emission,
and collision terms, focusing specifically on neutrino transfer in
high-density and high-temperature medium.

The paper is organized as follows: In \S~2, we review the covariant
moment formalism derived by Thorne \cite{Kip}. In \S~3, a truncated
moment formalism is presented, assuming a physically reasonable
specific form for the distribution function. In \S~4, source terms of
the moment formalism are written in terms only of the radiation
field variables employed in our truncated moment formalism, focusing
specifically on neutrino transfer. In \S~5, approximate solutions for
the radiation fields in the optically thick limit are derived. In
\S~6, we propose a closure relation among the radiation scalar,
vector, and tensor. We also derive the characteristic propagation
speeds of the radiation field in the optically thick and thin limits 
for the chosen closure relation. 
In \S~7, hydrodynamic equations coupled with the radiation fields are
derived. In \S~8, radiation hydrodynamic equations in a slow-motion
limit (usually referred to as Newtonian radiation hydrodynamic
equations) are derived. Section 9 is devoted to a summary.  Throughout
this paper, Greek ($\alpha$, $\beta$, $\gamma \cdots$) and Latin ($i$,
$j$, $k \cdots$) subscripts denote the spacetime and space components,
except for $\nu$ which always denotes the angular frequency of
radiation (which never be the subscript of space or time).  $x^{\mu}$
always denotes spacetime coordinates. We assume to use the Cartesian
coordinates as the spatial coordinates $x^i$ for simplicity.  Unless
otherwise stated, the units of $c=1=h$ are used, where $c$ is the speed
of light and $h$ the Planck constant. $k_{\rm b}$ denotes the
Boltzmann constant.

\section{Moment formalism of Thorne}

First, we review the Thorne's moment formalism~\cite{Kip}. In the
first step, he defines an unprojected moment of massless particles
associated with a moving medium as
\beqn
M_{(\nu)}^{~\alpha_1\alpha_2 \cdots \alpha_k}(x^{\beta})
=\int {f(p'^{\alpha},x^{\beta}) \delta(\nu-\nu') \over \nu'^{k-2}}
p'^{\alpha_1} p'^{\alpha_2} \cdots p'^{\alpha_k} dV'_p, \label{eq00}
\eeqn
%%%%%%%%%%%
where $f$ is the distribution function of the relevant radiation,
$\nu'=-u_{\mu}p'^{\mu}$ the frequency of the radiation in the
rest-frame of the medium (i.e, in the rest-frame of the fiducial
observer) with $u^{\mu}$ being medium's four velocity, $p^{\mu}$ the
four-momentum of the radiation, and $dV_p$ the invariant integration
element on the light cone.  $k$, here, is positive integer, 1, 2,
$\cdots$. As pointed out by Thorne~\cite{Kip}, the choice of the
fiducial observer is crucial when deriving a good truncated formalism from
his moment formalism. In the following, the fluid, coupled with the
radiation, is chosen as the medium~\cite{Kip,Lindquist,Castor}.
Namely, the frequency, $\nu$, in $M_{(\nu)}^{~\alpha_1\alpha_2 \cdots
\alpha_k}$ always denote the frequency {\em measured in the rest-frame of
the fluid} throughout this paper. This choice is crucially helpful 
when computing the source terms of the radiation equations.

We note that it is possible to choose any fiducial frame in the
moment formalism. However, we have to keep in mind that for a
truncated moment formalism in a closed form, it is necessary to
assume a closure relation which is determined by a physically
reasonable assumption. In the dense medium, radiation is strongly
coupled to the matter field. This implies that at the zeroth order,
the radiation is in equilibrium with the medium, and radiation flow 
(measured by an observer comoving with the matter) is a small
correction.  To reproduce this feature in the closure relation, the 
best method seems to choose the fluid rest frame as the fiducial frame. 

We also note the following: As a result of our choice of the fiducial
frame, the argument frequency in the distribution function is always
{\em the frequency measured in the fluid rest frame}. By contrast, the
argument variable should be in general the frequency in the laboratory
frame (although any frame can be taken), if one fully solves the
Boltzmann equation that the distribution function obeys.

The Boltzmann equation is written in the form~\cite{Kip}
\beqn
{dx^{\alpha} \over d\tau}{\pa f \over \pa x^{\alpha}}+{d p^{i} \over d\tau}
{\pa f \over \pa p^{i}}=(-p^{\alpha} u_{\alpha})S(p^{\mu}, x^{\mu}, f),
\eeqn
where $S$ denotes a source term and $\tau$ the affine parameter 
of a trajectory of radiation particles. In any orthonormal frame, 
the invariant integration element is given by \cite{Lindquist}
\beqn
dV_p={d\hat p^1 d\hat p^2 d\hat p^3 \over \hat p^0},
\eeqn
where $\hat p^{\alpha}$ is the four-momentum of the radiation 
in the local orthonormal frame. In the local rest frame of an 
observer comoving with the fluid, 
\beqn
dV_p=\nu d\nu d\Omega,
\eeqn
where $\int d\Omega$ denotes integrations over solid angle on 
an unit sphere. 

For the following, we write $p^{\alpha}$ in the form 
\beqn
{d x^{\alpha} \over d\tau}=p^{\alpha}=\nu (u^{\alpha} + \ell^{\alpha}), 
\eeqn
where $\ell^{\alpha}$ is a unit normal four-vector orthogonal to
$u^{\alpha}$;  $\ell_{\alpha} \ell^{\alpha}=1$ and 
$u_{\alpha}\ell^{\alpha}=0$. 
Using this decomposition of $p^{\alpha}$, Eq.~(\ref{eq00}) is
rewritten to give 
\beqn
M_{(\nu)}^{~\alpha_1\alpha_2 \cdots \alpha_k}
=\nu^3 \int f(\nu,\Omega,x^{\mu}) 
(u^{\alpha_1}+\ell^{\alpha_1})(u^{\alpha_2}+\ell^{\alpha_2})
\cdots (u^{\alpha_k}+\ell^{\alpha_k}) d\Omega \label{eq001}. 
\eeqn
Here, the angular dependence is included in $\ell^{\alpha}$ and 
the following relations hold, 
\beqn
&& \int d\Omega \ell^{\alpha}=0=
\int d\Omega \ell^{\alpha}\ell^{\beta}\ell^{\gamma},~~~~~
{1\over 4\pi}
\int d\Omega \ell^{\alpha} \ell^{\beta}={1\over 3}h^{\alpha\beta},
\nonumber \\
&& {1\over 4\pi}
\int d\Omega \ell^{\alpha} \ell^{\beta} \ell^{\gamma} \ell^{\delta}
={1\over 15}\Big(
h^{\alpha\beta}h^{\gamma\delta}
+h^{\alpha\gamma}h^{\beta\delta}
+h^{\alpha\delta}h^{\beta\gamma}
\Big).\label{angleinteg}
\eeqn
$h_{\alpha\beta}$ is the projection operator defined by 
\beqn
h_{\alpha\beta}:=g_{\alpha\beta}+u_{\alpha}u_{\beta}. 
\eeqn
Following Thorne~\cite{Kip}, we denote $M_{(\nu)}^{~\alpha_1 \alpha_2
\cdots \alpha_k}$ by $M_{(\nu)}^{~A_k}$. Taking the covariant
derivatives of $M_{(\nu)}^{~A_k \beta}$, we obtain a covariant 
equation with respect to real-space coordinates~\cite{Kip}
\beqn
\nabla_{\beta} M_{(\nu)}^{~A_k\beta}
-{\pa \over \pa \nu}(\nu M_{(\nu)}^{~A_k \beta\gamma} 
\nabla_{\gamma} u_{\beta})
-(k-1)M_{(\nu)}^{~A_k\beta\gamma}\nabla_{\gamma} u_{\beta}=S_{(\nu)}^{~A_k},
\label{eq002}
\eeqn
where $\nabla_{\alpha}$ denotes the covariant derivative 
associated with the spacetime metric $g_{\alpha\beta}$, and 
\beqn
S_{(\nu)}^{~A_k}=\nu^3 \int S(\nu, \Omega, x^{\mu}, f)
(u^{\alpha_1}+\ell^{\alpha_1})(u^{\alpha_2}+\ell^{\alpha_2})
\cdots (u^{\alpha_k}+\ell^{\alpha_k}) d\Omega \label{eq003}. 
\eeqn
%%%%%%%%%%%%
Here, the spacetime derivative is taken holding $\nu$ and the frequency
derivative is taken holding spacetime location.  It should be noted
that Eq.~(\ref{eq002}) has a coordinate-independent form as stressed by
Thorne~\cite{Kip}.  Also, the following relation is worthy to note:
%%%%%%%%%%%%
\beqn
M_{(\nu)}^{~A_k \beta}u_{\beta}=-M_{(\nu)}^{~A_k}. 
\eeqn
Thus, the rank-$(k+1)$ equations include the lower-rank equations. 

Since the frequency, $\nu$, in Eq.~(\ref{eq002}) denotes the frequency
{\em observed in a fluid-rest frame}, not in the laboratory frame,
$M_{(\nu)}^{~A_k}$ is not directly related to the spectrum observed in
the laboratory frame.  However, if the fluid is assumed to be at rest
in a distant zone far away from a radiation source where we observe
the spectrum, the radiation moments in the fluid-rest frame agree with
those in the laboratory frame. We suppose that the present formalism
will be used for the system that this assumption holds, e.g.,
supernova stellar core collapse and merger of compact objects. Thus,
it is possible to directly compute the radiation spectrum from
$M_{(\nu)}^{~A_k}$, if we estimate it for $r \rightarrow
\infty$ (cf. Appendix A for an example).

Integrating Eq.~(\ref{eq002}) by $\nu$, we obtain 
(for each species of the radiation component)
\beqn
\nabla_{\beta} M^{A_k\beta}
-(k-1)M^{A_k\beta\gamma}\nabla_{\gamma} u_{\beta}=S^{A_k},
\label{eq004}
\eeqn
where 
\beqn
M^{A_k}=\int_0^{\infty} d\nu M_{(\nu)}^{~A_k}~~{\rm and}~~
S^{A_k}=\int_0^{\infty} d\nu S_{(\nu)}^{~A_k}.
\eeqn
Equation (\ref{eq004}) is essentially the same as the moment 
formalism derived by Anderson and Spiegel \cite{AS}. We 
note that the second-rank tensor $M^{\alpha\beta}$ is equal to the 
energy-momentum tensor for one of the radiation components. 

In the following, we analyze only the second-rank part of
Eq.~(\ref{eq002}), truncating the higher-rank parts (in \S~5, we
partly use the third-rank equation for deriving a solution in the absence
of closure relation).  In the next section, we develop such formalism.

\section{Truncated moment formalism}\label{sec:trun}

%\subsection{Radiation fields}

First of all, we define the following moments: 
\beqn
&&J_{(\nu)}:= \nu^3 \int f(\nu,\Omega,x^{\mu}) d\Omega,\\
&&H_{(\nu)}^{~\alpha}:= \nu^3 \int \ell^{\alpha} f(\nu,\Omega,x^{\mu}) 
d\Omega,\\
&&L_{(\nu)}^{~\alpha\beta}:= \nu^3 \int \ell^{\alpha}\ell^{\beta} 
f(\nu,\Omega,x^{\mu}) d\Omega,\\
&&N_{(\nu)}^{~\alpha\beta\gamma}:= \nu^3 
\int \ell^{\alpha}\ell^{\beta} \ell^{\gamma} f(\nu,\Omega,x^{\mu}) 
d\Omega. 
\eeqn
Here, all these integrals are assumed to be performed in the 
local rest frame comoving with the fluid, and $\nu$ 
denotes the angular frequency of radiation measured in 
this local rest frame. The second- and third-rank moments are denoted by  
\beqn
&&M_{(\nu)}^{~\alpha\beta}=J_{(\nu)} u^{\alpha} u^{\beta} 
+ H_{(\nu)}^{~\alpha} u^{\beta} + H_{(\nu)}^{~\beta} u^{\alpha}
+L_{(\nu)}^{~\alpha\beta}, \label{eq10} \\
&&M_{(\nu)}^{~\alpha\beta\gamma}=J_{(\nu)} u^{\alpha} u^{\beta} u^{\gamma} 
+ H_{(\nu)}^{~\alpha} u^{\beta} u^{\gamma}
+ H_{(\nu)}^{~\beta}  u^{\alpha} u^{\gamma}
+ H_{(\nu)}^{~\gamma}  u^{\alpha} u^{\beta} \nonumber \\
&&~~~~~~~~~~+L_{(\nu)}^{~\alpha\beta} u^{\gamma}
+L_{(\nu)}^{~\alpha\gamma} u^{\beta}
+L_{(\nu)}^{~\beta\gamma} u^{\alpha}
+N_{(\nu)}^{~\alpha\beta\gamma}, \label{eq11} 
\eeqn
and the total stress-energy tensor for the radiation is 
\beqn
T_{\rm rad}^{\alpha\beta}=\sum \int_0^{\infty} d\nu M_{(\nu)}^{~\alpha\beta}, 
\eeqn
where the summation denotes to sum up for all the species of the radiation. 

In our truncated formalism, (i) we formally define the zeroth-,
first-, second- and third-rank moments from the distribution
function, and (ii) we solve the evolution equations only for the
zeroth- and first-rank moments. For the optically thick region, this
is approximately equivalent to assuming that the degree of anisotropy
of the distribution function in the fluid local rest frame is weak and
that the distribution function is approximated by 
\beqn
f(\nu,\Omega,x^{\mu})=f_0(\nu,x^{\mu})
+f_1^{\alpha}(\nu,x^{\mu})\ell_{\alpha}
+f_2^{\alpha\beta}(\nu,x^{\mu})\ell_{\alpha}\ell_{\beta}. 
\label{expand0}
\eeqn
Here, $f_0$, $f_1^{\alpha}$, and $f_2^{\alpha\beta}$ do not depend on the 
propagation angle of radiation in the fluid local rest frame
and $f_2^{\alpha\beta}$ is a traceless tensor with respect to 
$h_{\alpha\beta}$ (i.e., $f_2^{\alpha\beta}h_{\alpha\beta}=0$). 
We assume that $|f_0|$ is much larger than the absolute magnitude 
of $f_1^{\alpha}$ and $f_2^{\alpha\beta}$. For the expansion of 
Eq.~(\ref{expand0}), we obtain 
\beqn
&&J_{(\nu)}=4\pi \nu^3 f_0,\label{eq314}\\ 
&&H_{(\nu)}^{~\alpha}={4\pi \over 3} \nu^3 f_1^{\alpha},\\ 
&&L_{(\nu)}^{~\alpha\beta}={4\pi \over 3} \nu^3 \Big(f_0 h^{\alpha\beta} 
+{2\over 5}f_2^{\alpha\beta}\Big)={1\over 3} J_{(\nu)}h^{\alpha\beta}
+{8\pi \over 15} \nu^3 f_2^{\alpha\beta},\\
&&N_{(\nu)}^{~\alpha\beta\gamma}={1\over 5}
\Big(
H_{(\nu)}^{~\alpha} h^{\beta\gamma}+H_{(\nu)}^{~\beta} h^{\alpha\gamma}
+H_{(\nu)}^{~\gamma} h^{\alpha\beta}
\Big), \label{Nijk}
\eeqn
where we used the relations (\ref{angleinteg}).  Thus, $f_0$ and
$f_1^{\alpha}$ are directly related to $J_{(\nu)}$ and
$H_{(\nu)}^{~\alpha}$, and $f_2^{\alpha\beta}$ to the traceless part of
$L_{(\nu)}^{~\alpha\beta}$, respectively.  Because of the truncated
expansion for $f(\nu,\Omega,x^{\mu})$, we naturally obtain a closure
relation for $N_{(\nu)}^{~\alpha\beta\gamma}$. 

For the optically thin limit, by contrast, we should first give a
physical assumption {\em in the laboratory frame} because the
radiation does not interact with matter. We employ the assumptions
that the radiation should propagate with the speed of light and the
radiation flow at each spacetime point should be pointed to a null
direction. The former assumption then implies that the radiation
flow is pointed to a null direction in any frame (although the spacetime
coordinate basis changes). Thus, for such region, the distribution
function may be written in the form (see \S \ref{sec6.1} for details)
%%%%%%%%%%%%%%%%%%%
\beq
f(\nu,\Omega,x^{\mu})=4\pi f_{\rm f}(\nu,x^{\mu}) 
\delta(\Omega-\Omega_{\rm f}),\label{eq3.13}
\eeq
%%%%%%%%%%%%%%%%
where $\Omega_{\rm f}$ denotes the flow direction in the fluid rest
frame, and $f_{\rm f}(\nu,x^{\mu})$ is the partial distribution
function of $\Omega=\Omega_{\rm f}$. Then, the radiation moments are
calculated to give
%%%%%%%%%%%%%%%%
\beqn
&&J_{(\nu)}=4\pi \nu^3 f_{\rm f},\label{thin11}\\ 
&&H_{(\nu)}^{~\alpha}=4\pi \nu^3 f_{\rm f} \ell_{\rm f}^{\alpha},\\ 
&&L_{(\nu)}^{~\alpha\beta}=4\pi \nu^3 f_{\rm f} \ell_{\rm f}^{\alpha} 
\ell_{\rm f}^{\beta},\label{eq:lab}\\
&&N_{(\nu)}^{~\alpha\beta\gamma}=4\pi \nu^3 f_{\rm f} 
\ell_{\rm f}^{\alpha} \ell_{\rm f}^{\beta} \ell_{\rm f}^{\gamma},
\label{thin00}
\eeqn
%%%%%%%%%%%%%%%%%%
where $\ell_{\rm f}^{\alpha}$ denotes the unit vector of the flow
direction (observed in the fluid-rest frame). 
In Appendix A, we illustrate that the assumption of
(\ref{thin11})--(\ref{thin00}) would be appropriate for providing the
radiation field solution in the optically thin-limit medium. 

The equations for $J_{(\nu)}$ and $H_{(\nu)}^{~\alpha}$ are 
derived from the second-rank part of Eq.~(\ref{eq002}) as
\beqn
\nabla_{\beta} M_{(\nu)}^{~\alpha\beta}
-{\pa \over \pa \nu}
(\nu M_{(\nu)}^{~\alpha\beta\gamma} \nabla_{\gamma} u_{\beta})
=S_{(\nu)}^{~\alpha}, \label{eq012}
\eeqn 
where
\beqn
M_{(\nu)}^{~\alpha\beta\gamma} \nabla_{\beta} u_{\gamma}
&=&(H_{(\nu)}^{~\gamma} u^{\alpha}u^{\beta} +L_{(\nu)}^{~\alpha\gamma}
u^{\beta} +L_{(\nu)}^{~\beta\gamma} u^{\alpha} +
N_{(\nu)}^{~\alpha\beta\gamma} )\nabla_{\beta} u_{\gamma} \nonumber \\
&=&\Big(H_{(\nu)}^{~\gamma} u^{\alpha}+L_{(\nu)}^{~\alpha\gamma}\Big)
a_{\gamma} + \Big(L_{(\nu)}^{~\beta\gamma} u^{\alpha} +
N_{(\nu)}^{~\alpha\beta\gamma} \Big) \Sigma_{\beta\gamma}. 
\eeqn
%%%%%%%%%%
The acceleration $a^{\alpha}$ and the shear
$\Sigma_{\alpha\beta}$ are defined by 
\beqn
&&a^{\alpha}:=u^{\beta} \nabla_{\beta} u^{\alpha},\\
&&\Sigma_{\alpha\beta}:={1\over 2}
h_{\alpha}^{~\gamma} h_{\beta}^{~\delta}
\Big[\nabla_{\gamma} u_{\delta}+\nabla_{\delta} u_{\gamma}\Big]. 
\eeqn
To obtain a closed set of the equations, we have to determine
$L_{(\nu)}^{~\alpha\beta}$ [$N_{(\nu)}^{~\alpha\beta\gamma}$ is given
by Eq.~(\ref{Nijk}) or (\ref{thin00})].  Instead of solving the
equation for this, which may be derived from the moment equation of
third rank, we will assume a closure relation for it; an artificial
(but physically reasonable) relation between
$L_{(\nu)}^{~\alpha\beta}$ and $(J_{(\nu)}, H_{(\nu)}^{~\alpha})$ will
be assumed (see \S \ref{sec:clos}).

Substituting Eq.~(\ref{eq10}) into Eq.~(\ref{eq012}), the 
evolution equations for $J_{(\nu)}$ and $H_{(\nu)}^{~\alpha}$ are obtained as
\beqn
&&\nabla_{\alpha} Q_{(\nu)}^{~\alpha} 
+ Q_{(\nu)}^{~\alpha\beta} \nabla_{\beta} u_{\alpha}
- {\pa \over \pa \nu}[\nu (Q_{(\nu)}^{~\alpha\beta} \nabla_{\beta}
u_{\alpha})]
=-S_{(\nu)}^{~\alpha} u_{\alpha},\label{eqQ0}\\
%%%%%%%%%%%
&&h_{k\alpha}\Big[\nabla_{\beta} Q_{(\nu)}^{~\alpha\beta}
+Q_{(\nu)}^{~\beta} \nabla_{\beta} u^{\alpha}
%%-u^k Q_{(\nu)}^{~\alpha\beta} \nabla_{\beta} u_{\alpha}
- {\pa \over \pa \nu}[\nu
(L_{(\nu)}^{~\alpha\gamma}u^{\beta}+N_{(\nu)}^{~\alpha\beta\gamma})
\nabla_{\beta}u_{\gamma}]\Big]
=h_{k\alpha}S_{(\nu)}^{~\alpha},\label{eqQ1}
\eeqn
where
\beqn
&&Q_{(\nu)}^{~\alpha}:=
-M_{(\nu)}^{~\alpha\beta} u_{\beta}=J_{(\nu)} u^{\alpha} 
+ H_{(\nu)}^{~\alpha},\\
&&Q_{(\nu)}^{~\alpha\beta}:=h^{\alpha}_{~\gamma} M_{(\nu)}^{~\gamma\beta}=
H_{(\nu)}^{~\alpha} u^{\beta} + L_{(\nu)}^{~\alpha\beta}.
\eeqn
The frequency-integrated equations are
\beqn
&&\nabla_{\alpha} Q^{\alpha} 
+ Q^{\alpha\beta} \nabla_{\beta} u_{\alpha}
=-S^{\alpha} u_{\alpha},\\
%%%%%%%%%%%%%%%%%%%%%%%%%%%%%%%
&&h_{k\alpha} (\nabla_{\beta} Q^{\alpha\beta}
+Q^{\beta} \nabla_{\beta} u^{\alpha})
%%%-u^k Q^{\alpha\beta} \nabla_{\beta} u_{\alpha}
 =h_{k\alpha} S^{\alpha},
\eeqn
where
\beqn
Q^{\alpha}:=\int_0^{\infty} d\nu Q_{(\nu)}^{~\alpha},~~~
Q^{\alpha\beta}:=\int_0^{\infty} d\nu Q_{(\nu)}^{~\alpha\beta}.
\eeqn
Thus, the equations are not in the conservation form; the reason is 
that $Q^0$ and $Q^{k0}$ are not conservative quantities even 
in the absence of the source terms. 

Instead of using Eq.~(\ref{eq10}), $M_{(\nu)}^{~\alpha\beta}$ 
may be written by
\beq
M_{(\nu)}^{~\alpha\beta}=E_{(\nu)} n^{\alpha} n^{\beta} 
+ F_{(\nu)}^{~\alpha} n^{\beta} + F_{(\nu)}^{~\beta} n^{\alpha}
+P_{(\nu)}^{~\alpha\beta}, \label{eq10a}
\eeq
%%%%%%%%%%%%
where $n^{\alpha}$ is a unit vector orthogonal to the spacelike
hypersurface. $E_{(\nu)}$, $F_{(\nu)}^{~\alpha}$, and
$P_{(\nu)}^{~\alpha\beta}$ may be regarded as radiation fields measured
in the laboratory frame.  We note again that the meaning of the
frequency, $\nu$, is unchanged; it is the frequency observed in the
{\em fluid rest frame}. To obtain the quantities fully defined in the
laboratory frame, we need the transformation of $\nu$ to the frequency
measured in the laboratory frame. However, in the moment formalism, we
do not consider such transformation, as already mentioned.  

We however assume that $u^{\mu}=n^{\mu}$ in the far region with $r
\rightarrow \infty$ as mentioned in \S 2.  Our primary purpose is 
to develop an approximate formalism which can be used for simulation
of stellar core collapse and merger of binary compact objects. For
such purpose, this assumption is acceptable.  Thus, for $r \rightarrow
\infty$, we suppose that $E_{(\nu)}=J_{(\nu)}$,
$F_{(\nu)}^{~\alpha}=H_{(\nu)}^{~\alpha}$, and
$P_{(\nu)}^{~\alpha\beta}=L_{(\nu)}^{~\alpha\beta}$, and $\nu$ agrees
with the frequency measured in the laboratory frame. Therefore, if
$E_{\nu}$ is extracted in a distant zone in the numerical simulation,
we can obtain the spectrum of the radiation.

In the 3+1 formulation of general relativity,
%%%%%%%%%%
\beq
n^{\alpha}=\Big({1\over \alpha},-{\beta^k \over \alpha}\Big),
\eeq
where $\alpha$ is the lapse function and $\beta^k$ the shift 
vector. Then, $E_{(\nu)}$, $F_{(\nu)}^{~\alpha}$, and 
$P_{(\nu)}^{~\alpha\beta}$ are defined by
\beqn
E_{(\nu)}=M_{(\nu)}^{~\alpha\beta}n_{\alpha}n_{\beta},~~~
F_{(\nu)}^{~i}=-M_{(\nu)}^{~\alpha\beta}n_{\alpha}\gamma_{\beta}^{~i},~~~
P_{(\nu)}^{~ij}=M_{(\nu)}^{~\alpha\beta}
\gamma_{\alpha}^{~i}\gamma_{\beta}^{~j},
\eeqn
where $\gamma_{\alpha\beta}$ is the three metric
\beq
\gamma_{\alpha\beta}:=g_{\alpha\beta}+n_{\alpha}n_{\beta}.
\eeq
Because $F_{(\nu)}^{~\alpha}n_{\alpha}=P_{(\nu)}^{~\alpha\beta}n_{\alpha}=0$, 
we have the relations $F_{(\nu)}^{~0}=P_{(\nu)}^{~0\alpha}=0$. 

Here, we consider a formalism in which $E_{(\nu)}$ and $F_{(\nu)}^{~k}$
are evolved, and $P_{(\nu)}^{~ij}$ is determined by a closure
relation. Then, $J_{(\nu)}$ and $H_{(\nu)}^{~\alpha}$ are determined by
\beqn
&& J_{(\nu)}=E_{(\nu)} w^2 - 2F_{(\nu)}^{~k} w u_{k} 
+P_{(\nu)}^{~ij} u_{i} u_{j},\\
&& H_{(\nu)}^{~\alpha}=(E_{(\nu)}w -F_{(\nu)}^k u_k) 
h^{\alpha}_{~\beta} n^{\beta}
+w h^{\alpha}_{~\beta} F_{(\nu)}^{~\beta}
-h^{\alpha}_{~i} u_j P_{(\nu)}^{~ij}, \label{eq3.34}
%%&& H_{(\nu)}^{~\beta}=E_{(\nu)} w n^{\gamma} h_{\gamma}^{~\beta}
%%- F_{(\nu)}^{~\alpha} u_{\alpha} n^{\gamma} h_{\gamma}^{~\beta}
%%+ F_{(\nu)}^{~\alpha} h_{\alpha}^{~\beta} w 
%%- P_{(\nu)}^{~\alpha\gamma} u_{\alpha} h_{\gamma}^{~\beta},
\eeqn
%%%%%%%%
where $w=\alpha u^0$. We note $h^{\alpha}_{~\beta}
n^{\beta}=n^{\alpha}-w u^{\alpha}$ and $n_{\alpha}h^{\alpha\beta}
\gamma_{\beta k} =-w u_k$. For the later convenience, we give 
relations
\beqn
&&Q_{(\nu)}^{~\alpha} n_{\alpha}=-J_{(\nu)}w + H_{(\nu)}^{~\alpha} n_{\alpha}
=-E_{(\nu)} w + F_{(\nu)}^{~k} u_k,\\
&&Q_{(\nu)}^{~\alpha} \gamma_{\alpha i}=J_{(\nu)}u_i + H_{(\nu)i}
=w F_{(\nu)i}-P_{(\nu)i}^{~~k} u_k. 
\eeqn
%%%%%%%%%%%%%%%
As mentioned before, it is natural to assume that $u_i=0$ ($w=1/\alpha
\approx 1$) for the distant zone far from the radiation source. Then,
both the asymptotic power spectrum densities, $E_{(\nu)}$ and
$J_{(\nu)}$, agree with each other, because the frequency $\nu$ agrees
with that in the laboratory frame.

The evolution equations for $E_{(\nu)}$ and $F_{(\nu)i}$ are written 
in the conservative forms as 
\beqn
&& \pa_t (\sqrt{\gamma} E_{(\nu)})
+\pa_j[\sqrt{\gamma} (\alpha F_{(\nu)}^{~j} -\beta^j E_{(\nu)})]
+{\pa \over \pa \nu}
\Big(\nu \alpha\sqrt{\gamma}
n_{\alpha}M_{(\nu)}^{~\alpha\beta\gamma}\nabla_{\gamma}u_{\beta}
\Big)
\nonumber \\
&&\hskip 2cm 
=\alpha\sqrt{\gamma}[P_{(\nu)}^{~ij}K_{ij}-F_{(\nu)}^{~j} \pa_j \ln \alpha 
- S_{(\nu)}^{~\alpha} n_{\alpha}
],\label{eq1.2a}\\
%%%%%%%%%%%%%%%%%%%%%%%%%%%%%%%%%%%%
&&\pa_t (\sqrt{\gamma} F_{(\nu)i})
+\pa_j[\sqrt{\gamma} (\alpha P_{(\nu)i}^{~j} -\beta^j F_{(\nu)i})]
-{\pa \over \pa \nu}
\Big(\nu \alpha\sqrt{\gamma}
\gamma_{i\alpha}M_{(\nu)}^{~\alpha\beta\gamma}\nabla_{\gamma}u_{\beta}
\Big) 
\nonumber \\
&&\hskip 2cm 
=\sqrt{\gamma}\Big[-E_{(\nu)} \pa_i \alpha + F_{(\nu)k}\pa_i \beta^k
+{\alpha \over 2}P_{(\nu)}^{~jk}\pa_i \gamma_{jk} + \alpha S_{(\nu)}^{~\alpha}
\gamma_{i\alpha} \Big],
\label{eq1.2b}
\eeqn
where $\gamma$ is the determinant of $\gamma_{ij}$ 
and $K_{ij}$ the extrinsic curvature. 

The frequency-integrated equations are
\beqn
&& \pa_t (\sqrt{\gamma} E)
+\pa_j[\sqrt{\gamma} (\alpha F^{j} -\beta^j E)]
\nonumber \\
&&\hskip 2cm 
=\alpha\sqrt{\gamma}[P^{ij}K_{ij}-F^{j} \pa_j \ln \alpha 
- S^{\alpha} n_{\alpha}],\label{eq1.3a}\\
%%%%%%%%%%%%%%%%%%%%%%%%%%%%%%%%%%%%
&&\pa_t (\sqrt{\gamma} F_{i})
+\pa_j[\sqrt{\gamma} (\alpha P_{i}^{~j} -\beta^j F_{i})]
\nonumber \\
&&\hskip 2cm 
=\sqrt{\gamma}\Big[-E \pa_i \alpha + F_{k}\pa_i \beta^k
+{\alpha \over 2}P^{jk}\pa_i \gamma_{jk} + \alpha S^{\alpha}
\gamma_{i\alpha} \Big],
\label{eq1.3b}
\eeqn
where
\beqn
E:=\int_0^{\infty} d\nu E_{(\nu)},~~~
F_j:=\int_0^{\infty} d\nu F_{(\nu)j},~~~
P^{ij}:=\int_0^{\infty} d\nu P_{(\nu)}^{~ij}. 
\eeqn
%%%%%%%%%%%
Equations (\ref{eq1.3a}) and (\ref{eq1.3b}) have fully conservative
forms, because $E$ and $F_i$ are the conservative quantities in the
absence of the source terms and gravitational fields. Thus in the 
numerical simulation, it will be better to adopt basic equations based
on Eqs.~(\ref{eq1.2a}) and (\ref{eq1.2b}). 

\section{Source terms}

The source terms for the second-rank radiation field equations,
$S_{(\nu)}^{~\alpha}$, have to be written in terms of the radiation
moments $(J_{(\nu)}, H_{(\nu)}^{~\alpha}, L_{(\nu)}^{~\alpha\beta})$. 
$S_{(\nu)}^{~\alpha}$ is formally written as
%%%%%%%%%%%%%%%%%%
\beqn
S_{(\nu)}^{~\alpha}=\nu^3 \int B_{(\nu)}(\Omega,x^{\mu}) 
(u^{\alpha}+\ell^{\alpha}) d\Omega, \label{sinteg}
\eeqn
%%%%%%%%%
where $B_{(\nu)}$ is the so-called collision integral.  In the
following, we assume that $S_{(\nu)}$ and $B_{(\nu)}$ are written as 
a function of the phase-space coordinate defined in the
local rest frame of the fluid. The real coordinate, $x^{\mu}$, 
is arbitrarily chosen. 

We derive the source terms focusing on the neutrino transfer in a
high-density and high-temperature medium. We show that under a
reasonable and often-used assumption (anisotropy of the collision 
integral is small), the source terms are totally written in terms of
$J_{(\nu)}$, $H_{(\nu)}^{~\alpha}$, and $L_{(\nu)}^{~\alpha\beta}$. 

\subsection{Neutrino absorption and emission}

First, we consider the absorption and emission of neutrinos by nucleons
and heavy nuclei such as $n + \nu_e \leftrightarrow p + e^-$, $p + \bar
\nu_e \leftrightarrow n + e^+$, and $(Z,A) + \nu_e \leftrightarrow (Z-1,A)
+ e^-$, where $n$, $p$, $e^{\mp}$, $\nu_e$ ($\bar \nu_e$), and $(Z,A)$
denote neutrons, protons, electrons (positrons), electron neutrinos
(anti neutrinos), and heavy nuclei, respectively.  For these cases,
the collision integral is written in the form \cite{Bruen,Rampp}
%%%%%%%%
\beqn
B_{(\nu)}=j_{(\nu)}[1-f(\nu,\Omega,x^{\mu})]-{f(\nu,\Omega,x^{\mu})
\over \lambda_{(\nu)}},
\label{abem}
\eeqn
where $j_{(\nu)}$ denotes the emissivity, $\lambda_{(\nu)}$ is the
neutrino absorption mean free path, and $f(\nu,\Omega,x^{\mu})$
denotes the distribution function of relevant neutrinos (in the
following, we omit the argument $x^{\mu}$ in $f$).  $j_{(\nu)}$ and
$\lambda_{(\nu)}$ are quantities independent of the neutrino
propagation angle, $\Omega$. 

The integral of Eq.~(\ref{sinteg}) is easily performed to give
\beqn
S_{(\nu)}^{~\alpha}&&=4\pi j_{(\nu)} \nu^3 u^{\alpha} -
\Big(J_{(\nu)} u^{\alpha} +H_{(\nu)}^{~\alpha}\Big)
(j_{(\nu)}+\lambda_{(\nu)}^{-1}) \nonumber \\
&&=(j_{(\nu)}+\lambda_{(\nu)}^{-1}) 
\Big[(J_{(\nu)}^{\rm eq} -J_{(\nu)})u^{\alpha}-H_{(\nu)}^{~\alpha}
\Big], \label{abem2}
\eeqn
where 
\beq
J_{(\nu)}^{\rm eq}:=4\pi \nu^3 {j_{(\nu)} \over j_{(\nu)}
+\lambda_{(\nu)}^{-1}}
=4\pi \nu^3 f^{\rm eq}(\nu), 
\eeq
and $f^{\rm eq}(\nu)$ is the equilibrium distribution function.  
For neutrinos (fermions),
\beqn
f^{\rm eq}(\nu)={1 \over e^{(h\nu-\mu_{\rm c})/k_{\rm b}T}+1},
\eeqn
where $\mu_{\rm c}$ and $T$ are the chemical potential and the temperature
for the corresponding species of neutrinos which are in thermal 
equilibrium with matter. For the following, we define the opacity as
\beq
\kappa_{(\nu)}:=j_{(\nu)} + \lambda_{(\nu)}^{-1}, 
\eeq
and thus, 
\beqn
S_{(\nu)}^{~\alpha}=\kappa_{(\nu)}
\Big[(J_{(\nu)}^{\rm eq} -J_{(\nu)})u^{\alpha}-H_{(\nu)}^{~\alpha}\Big].
\eeqn

\subsection{Neutrino-electron scattering}

Neutrinos are scattered by electrons, nucleons, and heavy nuclei. 
In the case of electron scattering, the collision integral is generally 
written as \cite{Bruen,Rampp}
\beqn
B_{(\nu)}=\int \nu'^2 d\nu' d\Omega' 
&&[f(\nu',\Omega')\{1-f(\nu,\Omega)\} R^{\rm in}(\nu,\nu',\omega) 
\nonumber \\
&&-f(\nu,\Omega)\{1-f(\nu',\Omega')\} R^{\rm out}(\nu,\nu',\omega)], 
\eeqn
where $\omega$ is the cosine of the scattering angle, and 
$R^{\rm in}$ and $R^{\rm out}$ are the scattering kernels. 
Following Refs.~\citen{Bruen,Rampp}, we approximate these kernels by 
taking the terms up to the linear order in $\omega$, i.e.,  
\beqn
&&R^{\rm in}(\nu,\nu',\omega)=R^{\rm in}_0(\nu,\nu') 
+ R^{\rm in}_1(\nu,\nu') \omega,\\
&&R^{\rm out}(\nu,\nu',\omega)=R^{\rm out}_0(\nu,\nu') 
+ R^{\rm out}_1(\nu,\nu') \omega.
\eeqn
$\omega$ is related to the angular part of the momentum-space
coordinates $\Omega=(\theta, \varphi)$ and $\Omega'=(\theta',
\varphi')$ of the ingoing and outgoing neutrinos by
\beqn
\omega=\cos\theta \cos\theta' + \sin\theta\sin\theta' \cos(\varphi-\varphi'),
\eeqn
and thus, we have the following relations
\beqn
\int \omega d\Omega =0=\int \omega \ell^{\alpha}\ell^{\beta}d\Omega,~~~
\int \omega \ell^{\alpha} d\Omega = {4\pi \over 3} \ell'^{\alpha}.
\eeqn
Consequently, the collision integral is written as
\beqn
B_{(\nu)}&&=4\pi \int \nu'^2 d\nu'  
\Big[\{1-f(\nu,\Omega)\}\Big\{f_0(\nu')R^{\rm in}_0(\nu,\nu') 
+ {1 \over 3} f_1^{\alpha}(\nu')\ell_{\alpha}
R^{\rm in}_1(\nu,\nu') \Big\} 
 \nonumber \\ &&
-f(\nu,\Omega)\Big\{ \{(1-f_0(\nu')\}R^{\rm out}_0(\nu,\nu')
-{1 \over 3}f_1^{\alpha}(\nu')\ell_{\alpha}
R^{\rm out}_1(\nu,\nu') \Big\} \Big],~~~~~
\eeqn
and the source term is 
\beqn
&&S_{(\nu)}^{~\alpha}=
\int {d\nu' \over \nu'} \Big[
\big\{(4\pi\nu^3 - J_{(\nu)})u^{\alpha}-H_{(\nu)}^{~\alpha} \big\}
J_{(\nu')}R_0^{\rm in}(\nu,\nu') 
\nonumber \\ && ~~~~~~~~~~~~~~~~~
+{H_{(\nu')}^{~\alpha} \over 3}
\Big\{(4\pi \nu^3-J_{(\nu)})R_1^{\rm in}(\nu,\nu') 
+ J_{(\nu)}R_1^{\rm out}(\nu,\nu') \Big\}
\nonumber \\ && ~~~~~~~~~~~~~~~~~
-(h_{\gamma\sigma}H_{(\nu)}^{~\gamma}H_{(\nu')}^{~\sigma} u^{\alpha} 
+\tilde L_{(\nu)\beta}^{~\alpha}H_{(\nu')}^{~\beta})
\{R_1^{\rm in}(\nu,\nu')-R_1^{\rm out}(\nu,\nu')\} 
\nonumber \\ && ~~~~~~~~~~~~~~~~~
-(J_{(\nu)} u^{\alpha}+H_{(\nu)}^{~\alpha})
(4\pi\nu'^3 - J_{(\nu')})R_0^{\rm out}(\nu,\nu')
\Big], 
\eeqn
%%%%%%%%%%%%%%%%
where $\tilde L_{(\nu)}^{~\alpha\beta}$ is the traceless part of 
$L_{(\nu)}^{~\alpha\beta}$: 
\beq
\tilde L_{(\nu)}^{~\alpha\beta}:=L_{(\nu)}^{~\alpha\beta}
-{1\over 3}J_{(\nu)}h^{\alpha\beta}.
\eeq
Because of the approximation in which the terms up to the linear order in
$\omega$ for the scattering kernel is taken in account, the source
term is totally written in terms of $J_{(\nu)}$, $H_{(\nu)}^{~\alpha}$,
and $L_{(\nu)}^{~\alpha\beta}$. 

\subsection{Pair production}

For the thermal neutrino pair production and pair annihilation, the
collision integral has the following form \cite{Bruen,Rampp}
%%%%%%%%%%%%
\beqn
B_{(\nu)}=\int \nu'^2 d\nu' d\Omega' 
&&\bigl[\{1-f(\nu,\Omega)\}\{1-\bar f(\nu',\Omega')\} 
R^{\rm pro}(\nu,\nu',\omega) \nonumber \\
&&~~-f(\nu,\Omega)\bar f(\nu',\Omega')R^{\rm ann}(\nu,\nu',\omega)\bigr],
\eeqn
where $f$ and $\bar f$ are the distribution functions of neutrinos and
anti-neutrinos, respectively, and $R^{\rm pro}$ and $R^{\rm ann}$ are
the integration kernels for pair production and annihilation,
respectively.  This integral can be performed in the same manner as in
the neutrino-electron scattering: For the expansion up to the linear order
in $\omega$,
\beqn
&&R^{\rm pro}(\nu,\nu',\omega)=R^{\rm pro}_0(\nu,\nu') 
+ R^{\rm pro}_1(\nu,\nu') \omega,\\
&&R^{\rm ann}(\nu,\nu',\omega)=R^{\rm ann}_0(\nu,\nu') 
+ R^{\rm ann}_1(\nu,\nu') \omega, 
\eeqn
we obtain 
\beqn
&&S_{(\nu)}^{~\alpha}=
\int {d\nu' \over \nu'} \Big[
-\{(J_{(\nu)}-4\pi \nu^3)u^{\alpha}+H_{(\nu)}^{~\alpha}\} 
(4\pi\nu'^3 - \bar J_{(\nu')}) R_0^{\rm pro}(\nu,\nu')
\nonumber \\ && ~~~~~~~~~~~~~~~~~
-{\bar H_{(\nu')}^{~\alpha} \over 3} \Big\{
(4\pi \nu^3 -J_{(\nu)})R_1^{\rm pro}(\nu,\nu')
+J_{(\nu)} R_1^{\rm ann}(\nu,\nu') \Big\}
\nonumber \\ && ~~~~~~~~~~~~~~~~~
+(h_{\gamma\sigma}H_{(\nu)}^{~\gamma} \bar H_{(\nu')}^{~\sigma} u^{\alpha} 
+\tilde L_{(\nu)\beta}^{~\alpha}\bar H_{(\nu')}^{~\beta})
[R_1^{\rm pro}(\nu,\nu')-R_1^{\rm ann}(\nu,\nu')] 
\nonumber \\ && ~~~~~~~~~~~~~~~~~
-(J_{(\nu)}u^{\alpha}
+H_{(\nu)}^{~\alpha}) \bar J_{(\nu')} R_0^{\rm ann}(\nu,\nu')
\Big], 
\eeqn
%%%%%%%%%%%
where the quantities with bar denote the radiation moments for
anti-neutrinos. Again, the source term is totally written in terms
of $J_{(\nu)}$, $H_{(\nu)}^{~\alpha}$, and $L_{(\nu)}^{~\alpha\beta}$. 

\subsection{Isoenergy neutrino scattering}

In the neutrino scattering with nucleons and heavy nuclei, the energy
exchange may be assumed to be zero. In such isoenergetic neutrino
scattering, the collision integral is written as \cite{Bruen,Rampp}
%%%%%%%%
\beqn
B_{(\nu)}&=&\nu^2 \int d\Omega' 
[\{1-f(\nu,\Omega)\}f(\nu,\Omega')
-f(\nu,\Omega)\{1-f(\nu,\Omega')\}]R^{\rm iso}(\nu,\omega) \nonumber \\
&=&\nu^2 \int d\Omega' 
[f(\nu,\Omega')-f(\nu,\Omega)]R^{\rm iso}(\nu,\omega),
\eeqn
%%%%%%%%%
where $\nu$ denotes the angular frequency of the ingoing and outgoing
neutrinos.  Following Refs.~\citen{Bruen,Rampp}, we again approximate
the kernel $R^{\rm iso}(\nu,\omega)$ by taking the terms up to the
linear order in $\omega$ as
%%%%%%%%%%
\beqn
R^{\rm iso}(\nu,\omega)=R_0^{\rm iso}(\nu)+\omega R_1^{\rm iso}(\nu).
\eeqn
Then, 
\beqn
B_{(\nu)}=4\pi \nu^2 \Big[
{1 \over 3} f_1^{\alpha} \ell_{\alpha} R_1^{\rm iso}(\nu)
-\Big((f_1^{\alpha}\ell_{\alpha} + f_2^{\alpha\beta}
\ell_{\alpha}\ell_{\beta})
R_0^{\rm iso}(\nu)\Big)\Big],
\eeqn
and thus, 
\beqn
S_{(\nu)}^{~\alpha}=-\kappa_{(\nu)}^{\rm iso} H_{(\nu)}^{~\alpha}, 
%S_{(\nu)}^{~\alpha}=4\pi \nu^2 H_{(\nu)}^{~\alpha} \Big[
%{1 \over 3} R_1^{\rm iso}(\nu)
%-R_0^{\rm iso}(\nu)\Big]. 
\label{isos}
\eeqn
where
\beqn
\kappa_{(\nu)}^{\rm iso}=4\pi \nu^2 \Big[R_0^{\rm iso}(\nu)-
{1 \over 3} R_1^{\rm iso}(\nu)\Big]. 
\eeqn
Therefore, the source term depends only on the radiation flux 
(not on $J_{(\nu)}$), as clearly shown in Ref.~\citen{Rampp} [cf. 
Eq.~(A.46) of Ref.~\citen{Rampp}]. 

\section{Optically thick limit}\label{sec:thick}

In this section, we derive a solution of the radiation moment
equations for the limit that the radiation is optically thick to
absorption and scattering with matter fields. For neutrinos, such a
limit may be realized only for a special phenomena, such as stellar
core collapse and merger of binary neutron stars, in which an extreme
state of high density and high temperature is likely to be
realized. However, it is quite useful to derive the solution for the
idealized situation and to confirm that the derived solution is
physical, for validating a new formalism.

In the following, we analyze the radiation moment equations derived
in the previous sections, taking into account only the neutrino
absorption, emission, and isoenergy scattering for simplicity, for
which the source terms of the collision integral are written by
Eqs.~(\ref{abem2}) and (\ref{isos}). Then, the source terms of
Eqs.~(\ref{eqQ0}) and (\ref{eqQ1}) are in the form 
%%%%%%%%%%%%%%%%%%%%%%%%%%%%
\beqn
&& -S_{(\nu)}^{~\alpha} u_{\alpha}=\kappa_{(\nu)}
(J_{(\nu)}^{\rm eq} - J_{(\nu)}),\\
&&h_{k\alpha} S_{(\nu)}^{~\alpha}=-\tilde \kappa_{(\nu)}H_{(\nu)k}, 
\eeqn
where 
\beq
\tilde \kappa_{(\nu)}=\kappa_{(\nu)} + \kappa_{(\nu)}^{\rm iso}. 
%%+4\pi \nu^2 
%%\Big(R^{\rm iso}_0(\nu)-{1\over 3}R^{\rm iso}_1(\nu)\Big). 
\eeq
We  define a mean free path as the inverse of the opacity 
\beqn
l_{(\nu)}:=\kappa_{(\nu)}^{-1},
\eeqn
which measures the optical thickness of neutrinos: In the optically
thick limit, $l_{(\nu)}$ is much smaller than a characteristic length
scale of system (e.g., a stellar radius or $\rho/|\ell^{\mu}
\nabla_{\mu} \rho|$), and the radiation fields may be expanded by
$l_{(\nu)}$ assuming that it is sufficiently small \cite{AS} (this is
the so-called Thomas approximation~\cite{AS,Kip}). Assuming that $\tilde
\kappa_{(\nu)}^{-1}$ is also of order $l_{(\nu)}$, we can expand the
radiation moments as 
\beqn
&&J_{(\nu)}=J_{(\nu)}^{\rm eq} + l_{(\nu)} J_{(\nu)}^{(1)} + O(l_{(\nu)}^2),\\
&&H_{(\nu)}^{~\alpha}=\tilde l_{(\nu)} H_{(\nu)}^{(1)\alpha} 
+ O(l_{(\nu)}^2),\\
&&L_{(\nu)}^{~\alpha\beta}={1\over 3} J_{(\nu)} h^{\alpha\beta} 
+ l_{(\nu)} L_{(\nu)}^{(1)\alpha\beta} + O(l_{(\nu)}^2),\\
%%&&N_{(\nu)}^{~\alpha\beta\gamma}=O(l_{(\nu)}),
&&N_{(\nu)}^{~\alpha\beta\gamma}={\tilde l_{(\nu)} \over 5}\Big(
H_{(\nu)}^{(1)\alpha} h^{\beta\gamma}+H_{(\nu)}^{(1)\beta} h^{\alpha\gamma}
+H_{(\nu)}^{(1)\gamma} h^{\alpha\beta}
\Big)+O(l_{(\nu)}^2),
\eeqn
%%%%%%%%%%%%%%
where $\tilde l_{(\nu)}=\tilde \kappa_{(\nu)}^{-1}$.  In the following, we
determine $J_{(\nu)}$, $H_{(\nu)}^{~\alpha}$, and $L_{(\nu)}^{~\alpha\beta}$
up to the first order in $l_{(\nu)}~(\tilde l_{(\nu)})$.  For this purpose
(specifically for deriving $L_{(\nu)}^{(1)\alpha\beta}$), it is
necessary to analyze not only the second-rank moment equations but also the
third-rank moment equations
\beqn
\nabla_{\gamma} M_{(\nu)}^{~\alpha\beta\gamma}
-M_{(\nu)}^{~\alpha\beta\gamma\delta}\nabla_{\delta} u_{\gamma}
-{\pa \over \pa \nu}(\nu M_{(\nu)}^{~\alpha\beta\gamma\delta}
\nabla_{\delta} u_{\gamma})=S_{(\nu)}^{~\alpha\beta}.
\eeqn
Thus, we have to take into account the fourth-rank moment as
\beqn
U_{(\nu)}^{~\alpha\beta\gamma\delta}
:=\nu^3 \int \ell^{\alpha}\ell^{\beta}\ell^{\gamma}\ell^{\delta} 
f(\nu,\Omega,x^{\mu}) d\Omega,
\eeqn
and expand up to the zeroth order in $l_{(\nu)}$ as
\beqn
U_{(\nu)}^{~\alpha\beta\gamma\delta}
={1 \over 15} J_{(\nu)}^{\rm eq} 
(h^{\alpha\beta}h^{\delta\gamma}
+h^{\alpha\delta}h^{\beta\gamma}
+h^{\alpha\gamma}h^{\beta\delta}) + O(l_{(\nu)}). 
\eeqn
In addition, we have to calculate the second-rank moment of the source term 
which is 
\beqn
S_{(\nu)}^{~\alpha\beta}=&&
{1 \over l_{(\nu)}}\Big[(J_{(\nu)}^{\rm eq}-J_{(\nu)})\Big(u^{\alpha}u^{\beta}
+{1\over 3}h^{\alpha\beta}\Big)
-{l_{(\nu)} \over \tilde l_{(\nu)}}
(H_{(\nu)}^{~\alpha} u^{\beta}+H_{(\nu)}^{~\beta} u^{\alpha})
\nonumber \\
&&~~~~~~-{l_{(\nu)} \over 2 \bar l_{(\nu)}}(
\tilde L_{(\nu)}^{~\alpha\gamma}h_{\gamma}^{~\beta}
+\tilde L_{(\nu)}^{~\beta\gamma}h_{\gamma}^{~\alpha}) \Big],
\eeqn
where $\bar l_{(\nu)}^{-1}=\bar \kappa_{(\nu)}=\kappa_{(\nu)}+4\pi\nu^2
R_0^{\rm iso}$.

The expanded solutions for the radiation moments are determined from
the expanded equations for Eqs.~(\ref{eqQ0}) and (\ref{eqQ1}) in each
order in $l_{(\nu)}$. Their zeroth-order equations give the first-order 
solutions
%%%%%%%%%%%%%%
\beqn
&&J_{(\nu)}^{(1)}=-{1\over 3}
\Big[3 u^{\alpha} \nabla_{\alpha} J_{(\nu)}^{\rm eq}
+\Big(4 J_{(\nu)}^{\rm eq} 
-{\pa \over \pa \nu}(\nu J_{(\nu)}^{\rm eq})\Big) 
\nabla_{\alpha} u^{\alpha}
\Big],\\
&&H_{(\nu)}^{(1)\alpha}=-{1\over 3} h^{\alpha\beta}
\Big[\nabla_{\beta} J_{(\nu)}^{\rm eq}
+ \Big(4 J_{(\nu)}^{\rm eq} 
-{\pa \over \pa \nu}(\nu J_{(\nu)}^{\rm eq}) \Big)
u^{\gamma}\nabla_{\gamma} u_{\beta}
\Big],
\eeqn
and $L_{(\nu)}^{(1)\alpha\beta}$ is derived from the zeroth-order 
part of the third-rank moment equation. Taking into account that 
this term is traceless and its component is perpendicular to
$u^{\alpha}$, we obtain
\beqn
L_{(\nu)}^{(1)\alpha\beta}=-{\bar l_{(\nu)} \over 15 l_{(\nu)}} 
\Big[4 J_{(\nu)}^{\rm eq}
-{\pa \over \pa \nu}(\nu J_{(\nu)}^{\rm eq})\Big]\sigma^{\alpha\beta},
\eeqn
where
\beqn
\sigma^{\alpha\beta}:= h^{\alpha\gamma} h^{\beta\delta}
\Big[\nabla_{\gamma} u_{\delta}+\nabla_{\delta}
u_{\gamma} 
-{2\over 3}h_{\gamma\delta}\nabla_{\sigma} u^{\sigma}\Big].
\eeqn
%%%%%%%
For the frequency-integrated case, all these solutions agree with
those in Ref.~\citen{AS}.  The physical meaning of the first-order
correction for the frequency-dependent equations (except for the
diffusion effect associated with the term $\nabla_{\alpha}
J_{(\nu)}^{\rm eq}$) is essentially the same as that for the
frequency-integrated case: The first-order corrections of $J_{(\nu)}$,
$H_{(\nu)}^{~\alpha}$, and $L_{(\nu)}^{~\alpha\beta}$ are associated with
the fluid expansion ($\Theta=\nabla_{\alpha} u^{\alpha}$), the fluid
acceleration ($a_{\beta}=u^{\alpha} \nabla_{\alpha} u_{\beta}$), and
the fluid shear ($\sigma_{\alpha\beta}$). 

We note that from the derived first-order solutions, the second-order
solutions for $J_{(\nu)}$ and $H_{(\nu)}^{~\alpha}$ are easily derived
from Eqs.~(\ref{eqQ0}) and (\ref{eqQ1}):
%%%%%%%%%%%%%%%%%%
\beqn
&&J_{(\nu)}^{(2)}=-\nabla_{\alpha} (J_{(\nu)}^{(1)} u^{\alpha})
-{\tilde l_{(\nu)} \over l_{(\nu)}} \nabla_{\alpha} H_{(\nu)}^{(1)\alpha}
\nonumber\\
&&~~~~~~~~~~
+\nu {\pa \over \pa \nu}
\Big[{1 \over 3} J_{(\nu)}^{(1)} \Theta
+{\tilde l_{(\nu)} \over l_{(\nu)}} H_{(\nu)}^{(1)\alpha} a_{\alpha}
+{1 \over 2}L_{(\nu)}^{(1)\alpha\beta} \sigma_{\alpha\beta}
\Big],\\
%%%%%%%%%%%%%%%%%%%%%%%%%%%%%%%%%
&&H_{(\nu)}^{(2)\alpha}=-h^{\alpha}_{~\gamma}
\Big[ \nabla_{\beta} 
\Big(
H_{(\nu)}^{(1)\gamma}u^{\beta}
+{l_{(\nu)} \over \tilde l_{(\nu)}} L_{(\nu)}^{(1)\beta\gamma} 
\Big)
+H_{(\nu)}^{(1)\beta} \nabla_{\beta}u^{\gamma}
+{l_{(\nu)} \over \tilde 3l_{(\nu)}} 
\Big(\nabla^{\gamma} J_{(\nu)}^{(1)}+4J_{(\nu)}^{(1)} a^{\gamma}
\Big) \nonumber \\
&& ~~~~~~~~~~
-{\pa \over \pa \nu}\Big\{ \nu 
\Big({l_{(\nu)} \over 3\tilde l_{(\nu)}} J_{(\nu)}^{(1)} a^{\gamma}
+{l_{(\nu)} \over \tilde l_{(\nu)}}L_{(\nu)}^{(1)\beta\gamma} a_{\beta}
+{1 \over 3}H_{(\nu)}^{(1)\gamma}\Theta 
+ {1\over 5}H_{(\nu)}^{(1)\beta} \sigma_{\beta}^{~\gamma})
\Big)
\Big\}
\Big],
\eeqn
%%%%%%%%%%%%%%%
where $J_{(\nu)}^{(2)}$ and $H_{(\nu)}^{(2)\alpha}$ are the coefficients
of $l_{(\nu)}^2$ and $\tilde l_{(\nu)}^2$, respectively.  For clarifying
the order and for simplicity, we here assume that $l_{(\nu)}$ and
$\tilde l_{(\nu)}$ are constants. 

It is interesting to note that for the frequency-integrated
case~\cite{AS}, the magnitude of the first-order terms is proportional
to $J^{\rm eq}$ where for each species of neutrinos
%%%%%%%%%%%%%
\beq
J^{\rm eq}=\int_0^{\infty} d\nu J_{(\nu)}^{\rm eq}, 
%%%={7 \over 8}a_r T^4, 
\eeq
%%%%%%%%%%%
%%%and $a_r$ is the radiation constant. 
For the frequency-dependent equations, the magnitude depends
universally on
%%%%%%%%%%%
\beqn
J_{(\nu)}^{\rm eq}-{1\over 4}{\pa \over \pa \nu}(\nu J_{(\nu)}^{\rm eq})
&=&-{4\pi \nu^4 \over 4}{\pa f^{\rm eq}(\nu) \over \pa \nu} \nonumber
\\
&=&{h \nu \over 4k_{\rm b}T}J_{(\nu)}^{\rm eq}(1-f^{\rm eq}(\nu)). 
\eeqn 
%%%%%%%%%%%%%
Because of the presence of a factor $1-f^{\rm eq}(\nu)$ and
$f^{\rm eq}(\nu)$, these first-order corrections play a role only by
neutrinos of energy around $\mu_{\rm c} - k_{\rm b}T \alt h\nu \alt
\mu_{\rm c} + k_{\rm b}T$, i.e., near the Fermi surface, if the 
corresponding species of neutrinos is degenerate.  This is a 
reasonable consequence and characteristic property for fermions. 
\footnote{We note that only electron neutrinos can be 
degenerate in general. For anti electron neutrinos, $\mu_{\rm c} < 0$
and for muon and tau neutrinos, $\mu_{\rm c}=0$, when these neutrinos
are in thermal equilibrium with matter.  Thus, these are not
degenerate in general.}

The first-order solutions for $J_{(\nu)}$, $H_{(\nu)}^{~\alpha}$, and 
$L_{(\nu)}^{~\alpha\beta}$ may be used to constitute a diffusion equation 
from which the first-order solutions are produced. 
Substituting the first-order solution into Eq.~(\ref{eqQ0})
with replacement of $J_{(\nu)}^{\rm eq}$ to $J_{(\nu)}$ gives 
\beqn
&&\nabla_{\alpha} (J_{(\nu)} u^{\alpha})
-{1 \over 3} \nabla_{\alpha} \Big[\tilde l_{(\nu)} 
\Big(h^{\alpha\beta}\nabla_{\beta} J_{(\nu)}
+ 4 \tilde J_{(\nu)} a^{\alpha}\Big)\Big] 
\nonumber \\
&&~~~~ - {\nu \over 3}{\pa \over \pa \nu}\Big[
J_{(\nu)} \Theta -\tilde l_{(\nu)}
\Big(a^{\alpha} \nabla_{\alpha} J_{(\nu)} +4 \tilde J_{(\nu)}
a^{\alpha}a_{\alpha}\Big)
-{8 \over 5}l_{(\nu)} \tilde J_{(\nu)} \sigma_{\alpha\beta}
\sigma^{\alpha\beta}\Big] \nonumber \\
&&~~~~=\kappa_{(\nu)}(J_{(\nu)}^{\rm eq} - J_{(\nu)}),
\eeqn
where
\beqn
\tilde J_{(\nu)}=J_{(\nu)} - {1 \over 4} {\pa \over \pa \nu}(\nu J_{(\nu)}). 
\eeqn
%%%%%%%%%%
Thus in general, the diffusion equation is modified by the
acceleration and shear motion of the fluid and by neutrinos, 
if we do not assume the slow motion of the fluid. 

We note that in an FLD approximation, the first-order solution 
for $H_{(\nu)}^{~\alpha}$ is modified as
\beqn
H_{(\nu)}^{~\alpha}=-{\tilde l_{(\nu)} \over 
3 + \tilde l_{(\nu)} J_{(\nu)}^{-1} u^{\gamma}\nabla_{\gamma} J_{(\nu)}
} h^{\alpha\beta}
\Big[\nabla_{\beta} J_{(\nu)}
+ \Big(4 J_{(\nu)}-{\pa \over \pa \nu}(\nu J_{(\nu)}) \Big)
u^{\gamma}\nabla_{\gamma} u_{\beta}
\Big],
\eeqn
and is then substituted in Eq.~(\ref{eqQ0}). With this prescription,
the equation for $J_{(\nu)}$ reduces to a wave equation with the
characteristic speed $\sim c$ for the case that $\tilde l_{(\nu)}$ is
much longer than a characteristic length scale of the system.

\section{Closure relations} \label{sec:clos}

In the truncated moment formalism derived in \S 3, we proposed to
solve the equations for $E_{(\nu)}$ and $F_{(\nu)}^{~i}$ but not to solve
that for $P_{(\nu)}^{~ij}$, which is assumed to be determined in terms
of $E_{(\nu)}$ and $F_{(\nu)}^{~i}$.  In this section, we propose a
physically reasonable closure relation.

\subsection{Optically thin case}\label{sec6.1}

In the limit that the optical depth is zero, the emission, absorption,
and scattering are negligible. When the source term of the radiation
field equations can be neglected, the radiation freely propagates, and
the radiation moments should obey a wave equation with no source. 

One example for such region is the asymptotically flat region, far
from the radiation source where curved spacetime effects as well as
hydrodynamic effects play a tiny role (e.g., we may consider $u^{\mu}
\approx n^{\mu}$, $J_{(\nu)} \approx E_{(\nu)}$, and
$H_{(\nu)}^{~\alpha} \approx F_{(\nu)}^{~\alpha}$ as already mentioned
in \S 2 and 3). Thus, any closure relation assumed has to satisfy at least
the equations in the flat spacetime.

For the flat spacetime, we obtain the equation for $F_{(\nu)}^{~j}$
from Eq. (\ref{eq1.2a})
%%%%%%%%%%%%%%%%%%
\beq
\pa_j (\sqrt{\eta} F_{(\nu)}^{~j})=0,\label{ficon}
\eeq
where $\eta$ is the determinant of the flat three metric $\eta_{ij}$.
This provides a reasonable solution of $F_{(\nu)}^{~j}$ for the spatial
infinity; for the spherically symmetric flow, $F_{(\nu)}^{~r} \propto
r^{-2}$, and for the plane symmetric flow, $F_{(\nu)}^i =$constant for the
flow direction. On the other hand, Eq. (\ref{eq1.2b}) gives
\beqn
\pa_k (\sqrt{\eta}P_{(\nu)~j}^{~k}) ={\sqrt{\eta} \over 2}P_{(\nu)}^{~ik}\pa_j
\eta_{ik}. \label{pijcon}
\eeqn

For an appropriate solution of $E_{(\nu)}$, the following closure
relation is the first candidate (and is that we finally choose):
%%%%%%%%%%%
\beqn
P_{(\nu)}^{~\alpha\beta}=E_{(\nu)} {F_{(\nu)}^{~\alpha} F_{(\nu)}^{~\beta} 
\over \gamma_{ij}F_{(\nu)}^{~i} F_{(\nu)}^{~j}}. \label{clos1}
\eeqn
%%%%%%%%%%
This choice satisfies Eq.~(\ref{pijcon}) in the asymptotically flat
region.  This is regarded as a general relativistic extension of the
so-called M1 closure \cite{Livermore,Gon} for the optically thin-limit
region.  In this case, $E_{(\nu)} \propto r^{-2}$ for the spherical
symmetric flow and $E_{(\nu)}=$constant for the plane symmetric flow,
and hence, the reasonable condition is guaranteed. Furthermore,
$P_{(\nu)}^{~jk}\gamma_{jk}$ is equal to $E_{(\nu)}$ in this
condition, guaranteeing the necessary condition for the radiation
fields, $g_{\alpha\beta}T_{\rm rad}^{\alpha\beta}=0$.

It should be noted that this choice {\em with no modification} may not
be accepted in general relativity, because two of the characteristic
speeds of the wave equations for $E_{(\nu)}$ and $F_{(\nu)}^{~k}$ may
exceed the speed of light (see~\S~\ref{sec:speed} and Appendix B). The
pure choice of this closure relation is allowed only for
\beqn
E_{(\nu)}=\sqrt{\gamma_{ij}F_{(\nu)}^{~i} F_{(\nu)}^{~j}}, \label{eq6.4}
\eeqn
%%%%%%%%%%%%%
in which the characteristic speed is guaranteed to be equal to the
speed of light. This is guaranteed in the optically thin limit.
However, for $E_{(\nu)} > \sqrt{\gamma_{ij}F_{(\nu)}^{~i}
F_{(\nu)}^{~j}}$ which may often happen in a not-completely free
streaming region, the characteristic speed may exceed the speed of
light. This implies that an appropriate modification in the grey
region is required in the choice of this closure relation to satisfy
Eq.~(\ref{eq6.4}) (see \S~\ref{sec6.3} for a candidate choice of
variable Eddington factor in the grey region and Appendix C for a
satisfactory test result).

Another possible candidate is
%%%%%%%%%%%%%%%%
\beqn
P_{(\nu)}^{~\alpha\beta}={F_{(\nu)}^{~\alpha} F_{(\nu)}^{~\beta} 
\over \sqrt{\gamma_{ij}F_{(\nu)}^{~i} F_{(\nu)}^{~j}}}. \label{clos2}
\eeqn
%%%%%%%%%%%%%%%%
This choice also satisfies Eq.~(\ref{pijcon}) in the asymptotic region.
With Eq.~(\ref{clos2}), Eq.~(\ref{pijcon}) for the spherical and
plane-symmetric stationary flows is written as
%%%%%%%%%%%%%%%%%
\beqn
\sqrt{\eta} F_{(\nu)}^{~k} \pa_k \hat n^{i}=0, \label{cond}
\eeqn
where $\hat n^k$ is a unit vector parallel to $F_{(\nu)}^{~k}$
\beq
\hat n^k:={F_{(\nu)}^{~k} \over \sqrt{\gamma_{ij}F_{(\nu)}^{~i} F_{(\nu)}^{~j}}}.
\eeq 
For the spherical and plane-symmetric stationary flows, 
the condition (\ref{cond}) is guaranteed, and hence, 
the closure relation (\ref{clos2}) is acceptable. 

With this choice, the characteristic speed of the equation for
$F_{(\nu) i}$ is approximately the speed of light in the
asymptotically flat region (see \S~\ref{sec:speed}). However,
$P_{(\nu)}^{~jk}\gamma_{jk}$ is not a priori guaranteed to be equal to
$E_{(\nu)}$; for this condition to be satisfied,
$\sqrt{\gamma_{ij}F_{(\nu)}^{~i} F_{(\nu)}^{~j}}$ has to be equal to
$E_{(\nu)}$ but the condition will not be satisfied in the near zone,
i.e., near the emission source (cf. Appendix B). Only in the far zone,
this condition seems to be followed from Eq.~(\ref{eq1.2a}) because
the radiation propagates with the speed of light as $|F_{(\nu)}^{~k}|
\sim E_{(\nu)}$. Because of this reason, we choose Eq.~(\ref{clos1})
for the closure relation. 

It is reasonable to suppose that Eqs.~(\ref{clos1}) and (\ref{eq6.4})
are satisfied in the optically thin limit. Because we have
$F_{(\nu)}^{\alpha}=E_{(\nu)} f^{\alpha}$ where $f^{\alpha}$ is a unit
spatial vector, $f_{\alpha}f^{\alpha}=1$, and orthogonal to 
$n^{\alpha}$, $n^{\alpha}f_{\alpha}=0$. $J_{(\nu)}$,
$H_{(\nu)}^{~\alpha}$, and $L_{(\nu)}^{~\alpha\beta}$ are rewritten by
%%%%%%%%%%%%%%%%%%%%%%%%%
\beqn
&&J_{(\nu)}=M_{(\nu)}^{~\alpha\beta}u_{\alpha}u_{\beta}
=E_{(\nu)}(u^{\alpha} q_{\alpha})^2,\\
&&H_{(\nu)}^{~\alpha}=-M_{(\nu)}^{~\beta\gamma}u_{\beta} h_{\gamma}^{~\alpha}
=-E_{(\nu)}u^{\beta} q_{\beta}h_{\gamma}^{~\alpha}q^{\gamma},\\
&&L_{(\nu)}^{~\alpha\beta}=M_{(\nu)}^{~\gamma\delta}
h_{\gamma}^{~\alpha}h_{\delta}^{~\beta}
=E_{(\nu)}h_{\gamma}^{~\alpha}h_{\delta}^{~\beta} q^{\gamma} q^{\delta}, 
\eeqn
%%%%%%%%%%%%%%%%%%
where $q^{\alpha}=n^{\alpha}+f^{\alpha}$ is a null vector. Defining
$4\pi \nu^3 f_{\rm f} \equiv E_{(\nu)} (u^{\alpha} q_{\alpha})^2$ and
$l_{\rm f}^{\alpha} \equiv -h_{\beta}^{~\alpha} q^{\beta}/(u^{\mu}
q_{\mu})$, we obtain Eqs.~(\ref{thin11})--(\ref{eq:lab}), and find it
reasonable to assume Eq.~(\ref{eq3.13}) as the distribution function
in the optically thin limit. It is easy to confirm that $l_{\rm
f}^{\alpha} l_{{\rm f}\alpha}=1$ for the above definition. 

Finally, we have to give a closure relation for the third-rank
moment. We propose to employ
%%%%%%%%%%
\beqn
N_{(\nu)}^{~\alpha\beta\gamma}=
{J_{(\nu)} H_{(\nu)}^{~\alpha} H_{(\nu)}^{~\beta} H_{(\nu)}^{~\gamma} 
\over (h_{\alpha\beta}H_{(\nu)}^{~\alpha} H_{(\nu)}^{~\beta})^{3/2}}, 
\label{clos3N1}
\eeqn
or
\beqn
N_{(\nu)}^{~\alpha\beta\gamma}=
{H_{(\nu)}^{~\alpha} H_{(\nu)}^{~\beta} H_{(\nu)}^{~\gamma} 
\over h_{\alpha\beta}H_{(\nu)}^{~\alpha} H_{(\nu)}^{~\beta}}.
\label{clos3N2}
\eeqn
Here $H_{(\nu)}^{~\alpha}$ is related to $E_{(\nu)}$, $F_{(\nu)}^{~i}$, 
and $P_{(\nu)}^{~ij}$ by Eq.~(\ref{eq3.34}). 
Associated with the choice of Eq.~(\ref{clos1}), we 
choose Eq.~(\ref{clos3N1}). 

\subsection{Optically thick case}

As shown in \S~\ref{sec:thick}, in the optically thick limit with
$l_{(\nu)} \rightarrow 0$ (or $\kappa_{(\nu)} \rightarrow \infty$),
$L_{(\nu)}^{~\alpha\beta}$ is written as
%%%%%%%%%%%
\beqn
L_{(\nu)}^{~\alpha\beta}={1 \over 3} h^{\alpha\beta} J_{(\nu)}
-{4 \over 15}\bar l_{(\nu)} \sigma^{\alpha\beta} 
\Big[J_{(\nu)}-{1\over 4}{\pa \over \pa \nu}(\nu J_{(\nu)})\Big]
+ O(l_{(\nu)}^2). \label{kj}
\eeqn
%%%%%%%%
%%Here, we define $l_{(\nu)}$ only taking into account the neutrino
%%absorption and emission.  
The correction term of $O(l_{(\nu)})$ is associated with the so-called
radiation viscosity. With this prescription, we can incorporate the
first-order effect of $L_{(\nu)}^{~\alpha\beta}$ in $l_{(\nu)}$ without
solving the third-rank moment equation. (Note that
$N_{(\nu)}^{~\alpha\beta\gamma}$ should be given by Eq.~(\ref{Nijk}) in
the present formalism.)  For the case that velocity of the medium is
much smaller than the speed of light, $|v^i| \ll c$, the term of
$O(l_{(\nu)})$ may be neglected. This term will play a role for the
medium moving around a black hole, such as in black hole accretion
disks.

For numerical computation, we have to transform the relation of
Eq.~(\ref{kj}) to the relation of $P_{(\nu)}^{~ij}$ as a function of
$E_{(\nu)}$ and $F_{(\nu)}^{~i}$. For this purpose, we omit the term,
$\pa (\nu J_{(\nu)})/\pa \nu$, for simplicity.  The reason is that in
its presence, $P_{(\nu)}^{~ij}$ is not written by $E_{(\nu)}$ and
$F_{(\nu)}^{~i}$ in a straightforward manner (although it is possible
to do in an approximate manner).  For the frequency-integrated case,
this term vanishes, and hence, we may say that the radiation viscosity
effect is taken into account in a frequency-averaged way. However, in
this treatment, the low-energy neutrinos, which should not contribute
to the radiation viscosity for degenerate neutrinos, may incorrectly
play a role. To avoid this unphysical contribution, it will be
appropriate to artificially reduce $\bar l_{(\nu)}$ to zero for $h\nu
\alt \mu_{\rm c}-k_{\rm b}T$ when treating degenerate neutrinos. 

Assuming that Eq. (\ref{kj}) holds with the omission of the 
third term, we have the relations
\beqn
&& E_{(\nu)}=\Big[{4w^2-1 \over 3} - \sigma_0\Big]J_{(\nu)} 
+2H_{(\nu) j} V^j,\label{eqE} \\
&& F_{(\nu)i}= \Big[{4 \over 3}w u_i +\sigma_i \Big]J_{(\nu)} 
+ w H_{(\nu)i} + {u_i \over w} H_{(\nu)j} V^j, \label{eqF}
\eeqn
where $V^i=\gamma^{ij}u_j~(V_i=u_i)$, and
\beqn
\sigma_0={4 \bar l_{(\nu)} \over 15} \sigma^{\alpha\beta} n_{\alpha}
n_{\beta}, ~~~
\sigma_i={4 \bar l_{(\nu)} \over 15} \sigma^{\alpha\beta} n_{\alpha}
\gamma_{\beta i}.
\eeqn
Also, we used $H_{(\nu)\alpha}u^{\alpha}=0$ and 
$H_{(\nu)}^{~0}=(\alpha w)^{-1}H_{(\nu)i} V^i$.
Equations (\ref{eqE}) and (\ref{eqF}) constitute simultaneous
equations for $J_{(\nu)}$ and $H_{(\nu)i}$. Inverting them yields 
\beqn
&& J_{(\nu)} =\Big[{2w^2+1 \over 3} + \sigma_0 \Big]^{-1} 
\biggl[(2w^2-1)E_{(\nu)} -2w F_{(\nu)}^{~k} u_k \biggr],\label{Jeq}\\
%%%%%%%%%%%%%%%%%%%%%
&& H_{(\nu)i}={1 \over w} F_{(\nu)i} 
+ {1 \over w(2w^2+1+3\sigma_0)}
\biggl[-[4w^3u_i + 3(2w^2-1)\sigma_i + 3\sigma_0 w u_i]E_{(\nu)}
\nonumber \\
&& \hskip 5.8cm
+[(4w^2+1)u_i+6w \sigma_i + 3\sigma_0 u_i]F_{(\nu)}^{~k} u_k \biggr]. 
\label{Heq}
\eeqn
Note that $F_{(\nu)}^{~k} u_k=F_{(\nu)k} V^k$ but $H_{(\nu)k} V^k
\not= H_{(\nu)}^{~k} u_k$; $H_{(\nu)}^{~k}=(\gamma^{kl}-\beta^k
\gamma^{lm}u_m/\alpha w)H_{(\nu)l}$. Also $w\sigma_0=-\sigma_i V^i$.
Then, $P_{(\nu)}^{~ij}$ is given by
\beqn
P_{(\nu)}^{~ij}&=&J_{(\nu)}
\Big[{\gamma^{ij}+4V^i V^j \over 3}
-{4 \bar l_{(\nu)} \over 15}\sigma^{kl} \gamma_k^{~i} \gamma_l^{~j}\Big]
+H_{(\nu)}^{~i} V^j + H_{(\nu)}^{~j} V^i, \label{eq616}
%%\nonumber \\
%%&=& {E_{(\nu)} \over 2w^2+1}\Big[(2w^2-1)\gamma^{ij}-4V^i V^j \Big]
%%+{1 \over w}(F_{(\nu)}^{~i} V^j + F_{(\nu)}^{~j} V^i) \nonumber \\
%%&&~+{2F_{(\nu)}^{~k} u_k \over (2w^2+1)w}(-w^2 \gamma^{ij}+V^i V^j).
\eeqn
%%%%%%%%%%%%%%
where $J_{(\nu)}$ and $H_{(\nu)}^{~k}(=\gamma^k_{~\mu}H_{(\nu)}^{~\mu}$) are
given by Eqs.~(\ref{Jeq}) and (\ref{Heq}). With this closure relation
for $P_{(\nu)}^{~ij}$, the necessary condition for the radiation fields,
$g_{\alpha\beta}T_{\rm rad}^{\alpha\beta}=0$, is guaranteed to be
satisfied.

We note that with the closure relation (\ref{eq616}), the first-order
term in $l_{(\nu)}$ may be accidentally larger than the zeroth-order
term for a high value of $\sigma_{ij}$. Thus, it may be necessary to
change the definition of $\bar l_{(\nu)}$ as
\beqn
\bar l_{(\nu)}={\rm min}\Big[{1\over \bar \kappa_{(\nu)}}, C_{\sigma}
\Big({V^ku_k \over \sigma^{\alpha\beta}\sigma_{\alpha\beta}}\Big)^{1/2}
\Big],
\eeqn
where $C_{\sigma}$ is a coefficient smaller than unity. 

\subsection{Grey zone}\label{sec6.3}

For a solution of the radiation fields in the optically grey zone, in
general, it is necessary to fully solve the radiation transfer
equation in general relativity. However, it is not possible in the
framework of truncated moment formalism and far beyond the scope of
this paper. We propose an approximate method which is essentially the
same as the variable Eddington factor method \cite{Livermore}. In this
prescription, $P_{(\nu)}^{ij}$ is given by 
\beqn
P_{(\nu)}^{~ij}={3\chi - 1\over 2}(P_{(\nu)}^{~ij})_{\rm thin}
%%{F_{(\nu)}^{~i} F_{(\nu)}^{~j} \over \sqrt{\gamma_{kl}F_{(\nu)}^{~k}
%%F_{(\nu)}^{~l}}}
+{3(1-\chi) \over 2}(P_{(\nu)}^{~ij})_{\rm thick}, \label{eqpij}
%%\biggl[
%%{E_{(\nu)} \over 2w^2+1}\Big[(2w^2-1)\gamma^{ij}-4V^i V^j \Big] \nonumber \\
%%&&~~~~~~~~~~~~~~~~+{1 \over w}(F_{(\nu)}^{~i} V^j + F_{(\nu)}^{~j} V^i) 
%%+{2F_{(\nu)}^{~k} u_k \over (2w^2+1)w}(-w^2 \gamma^{ij}+V^i V^j) \biggr],
\eeqn
where $\chi$ is the so-called variable Eddington factor, which is 
$\chi=1/3$ in the optically thick limit and $\chi=1$ in the 
optically thin limit. Following Ref.~\citen{Livermore}, we choose 
that $\chi$ is a function of $\bar F$, for which in general
relativity, the candidates are 
\beqn
{\bar F}:=\biggl( {\gamma_{ij} F_{(\nu)}^{~i}  F_{(\nu)}^{~j} 
\over E_{(\nu)}^2}\biggr)^{1/2}, \label{edd1}
\eeqn
and 
\beqn
{\bar F}:=\biggl( {h_{\alpha\beta} H_{(\nu)}^{~\alpha}  H_{(\nu)}^{~\beta} 
\over J_{(\nu)}^2}\biggr)^{1/2}. \label{edd2}
\eeqn
%%%%%%%%%%%%%%%%
For the optically thick and thin limits, ${\bar F}=0$ and ${\bar
F}=1$, respectively. For giving a correct value of ${\bar F}$ in the
optically thick limit, Eq.~(\ref{edd2}) should be chosen because
$H_{(\nu)}^{~\alpha}$ should be zero in the comoving frame; if the fluid
has a large uniform velocity, the value of ${\bar F}$ in
Eq.~(\ref{edd1}) would be highly different from zero even in an
optically thick medium. For giving a correct value of ${\bar F}$ in
the optically thin limit, both Eqs.~(\ref{edd1}) and (\ref{edd2}) can
be chosen, because in such a limit, $M_{(\nu)}^{~\alpha\beta}$ is 
proportional to $J_{(\nu)}p^{\alpha}p^{\beta}$ ($p^{\alpha}$ is 
a null vector) and ${\bar F}=1$ for the null fluid in both definitions 
(see \S 3). For this reason, we choose Eq.~(\ref{edd2}) for $\bar F$. 

With the choice of (\ref{edd2}), $\bar F$ obeys an algebraic equation 
for a given set of $E_{(\nu)}$ and $F_{(\nu)}^{~j}$. This can be written 
in the form
\beqn
\bar F^2 = {h_{\alpha\gamma}M_{(\nu)}^{~\alpha\beta}u_{\beta}
M_{(\nu)}^{~\gamma\sigma}u_{\sigma} \over 
M_{(\nu)}^{~\alpha\beta}u_{\alpha}u_{\beta}},
\eeqn
where for $M_{(\nu)}^{~\alpha\beta}$, Eq.~(\ref{eq10a}) 
is used with Eq.~(\ref{eqpij}). In numerical simulation, 
we have to solve this equation numerically. 

Livermore proposed several functions for $\chi({\bar F})$, e.g.,
\beqn
\chi={3+4{\bar F}^2 \over 5+2\sqrt{4-3{\bar F}^2}}. \label{chi1}
\eeqn
%%or 
%%\beqn
%%\chi=1-{2 \over Z} [\coth Z - Z^{-1}]~~{\rm or}~~
%%\chi=\coth Z [\coth Z - Z^{-1}],
%%\eeqn
%%%%%%%%%%
%%with ${\bar F}=\coth Z - Z^{-1}$. 
In the Appendix C, we employ Eq.~(\ref{chi1}), and show that it is
likely to work well.  However, it should be kept in mind that it might 
not be the best one and better closure relations should be further
explored.

For completeness, we have to provide $N_{(\nu)}^{~\alpha\beta\gamma}$. 
We propose to employ 
\beqn
N_{(\nu)}^{~\alpha\beta\gamma}
={3\chi - 1\over 2}(N_{(\nu)}^{~\alpha\beta\gamma})_{\rm thin}
+{3(1-\chi) \over 2}(N_{(\nu)}^{~\alpha\beta\gamma})_{\rm thick},
\eeqn
where $(N_{(\nu)}^{~\alpha\beta\gamma})_{\rm thin}$ 
and $(N_{(\nu)}^{~\alpha\beta\gamma})_{\rm thick}$ are 
given by Eqs.~(\ref{clos3N1}) and (\ref{Nijk}), respectively. 

\subsection{Characteristic speed}\label{sec:speed}

For numerical computation with conservation schemes, it is necessary
to know characteristic speeds.  Furthermore, the analysis of the
characteristic speed is helpful to check whether the proposed closure
relation is acceptable or not (i.e., it is smaller than the 
speed of light).

The characteristic speed of the radiation moment equations is computed
from the Jacobian matrix (e.g., Refs.~\citen{font,font2,S03}). For the
conservative variables $E_{(\nu)}$ and $F_{(\nu)}^{~i}$, the Jacobian
matrix for the $x$ direction is (in the following, we omit the
subscript $\nu$)
\beqn
A_{ab}=\left[
\begin{array}{llll}
-\beta^x & \alpha \gamma^{xx} & \alpha \gamma^{xy} & \alpha \gamma^{xz} \\
\displaystyle \alpha {\pa P^x_{~x} \over \pa E}
& \displaystyle -\beta^x+\alpha {\pa P^x_{~x} \over \pa F_x}
& \displaystyle \alpha {\pa P^x_{~x} \over \pa F_y}
& \displaystyle \alpha {\pa P^x_{~x} \over \pa F_z} \\
\displaystyle \alpha {\pa P^x_{~y} \over \pa E}
& \displaystyle \alpha {\pa P^x_{~y} \over \pa F_x}
& \displaystyle -\beta^x+\alpha {\pa P^x_{~y} \over \pa F_y}
& \displaystyle \alpha {\pa P^x_{~y} \over \pa F_z} \\
\displaystyle \alpha {\pa P^x_{~z} \over \pa E}
& \displaystyle \alpha {\pa P^x_{~z} \over \pa F_x}
& \displaystyle \alpha {\pa P^x_{~z} \over \pa F_y}
& \displaystyle -\beta^x+\alpha {\pa P^x_{~z} \over \pa F_z}
\end{array}
\right]. 
\eeqn
The characteristic speeds are computed from 
\beq
\det(A_{ab}-\lambda I_{ab})=0,  
\eeq
where $I_{ab}$ is the unit matrix.

For the optically thin
case with the closure relation (\ref{clos1}),
\beqn
\lambda=-\beta^x \pm \alpha {F^x \over \sqrt{F_k F^k}}~~{\rm and}~~
\lambda=-\beta^x+\alpha E {F^x \over F_k F^k} ~~({\rm double}),
\label{eq:chara1}
\eeqn
and with the closure relation (\ref{clos2}),
\beqn
\lambda=-\beta^x ~~{\rm and}~~
\lambda=-\beta^x+\alpha {F^x \over \sqrt{F_k F^k}} ~~({\rm triple}). 
\eeqn
%%%%%%%%%%%%%%%%%%%%%%%%%%%%%%%%%%%%%%%%%%%%%%%%%%%%%%%%%%%%%%
As we already pointed out, $|\lambda|$ can exceed the speed of light
for the closure relation (\ref{clos1}) if the opacity is not zero
limit; $|E F^x/F_k F^k|$ may exceed the speed of light. Hence, it is
not allowed to be employed without appropriate choice of the variable
Eddington factor for the optically grey zone. By contrast, with the
closure relation (\ref{clos2}), the characteristic speed is smaller
than the speed of light (but in this case, the tracefree condition for
the stress-energy tensor is not satisfied in general, as already
mentioned).

For the optically thick limit with $l_{(\nu)}=0$,
\beqn
&&\lambda=-\beta^x+{2 w^2 p \pm \sqrt{\alpha^2 \gamma^{xx}(2w^2+1)-2w^2 p^2}
\over 2w^2+1} \nonumber \\
&&{\rm and}~~\lambda=-\beta^x+p~~({\rm double}),\label{chara0}
\eeqn
where $p=\alpha V^x/w$. For $w=1~(p=0)$, the first one reduces to
\beqn
\lambda=-\beta^x  \pm \alpha \sqrt{{\gamma^{xx} \over 3}},
\eeqn
and thus, we obtain a well-known characteristic speed for the
radiation fluid in the diffusion limit ($\sim 1/\sqrt{3}$). For $w
\rightarrow \infty$ ($p \rightarrow \alpha/\sqrt{\gamma_{xx}} < 1$),
\beqn
\lambda \rightarrow -\beta^x +p,
\eeqn
and thus, $\lambda$ approaches to a local light speed (but never
exceeds it).  For $l_{(\nu)} \not=0$, the characteristic speed is
modified in a complicated form. However as far as $l_{(\nu)}$ is
small, this effect is not important.

The general formula for the closure relation (\ref{clos1}) is written
in the following manner. For $i$-direction,
\beqn
\lambda=-\beta^i \pm \alpha {F^i \over \sqrt{F_k F^k}}~~{\rm and}~~
\lambda=-\beta^i+\alpha E {F^i \over F_k F^k} ~~({\rm double}),
\eeqn
%%or 
%%\beqn
%%\lambda=-\beta^i ~~{\rm and}~~
%%\lambda=-\beta^i+\alpha {F^i \over \sqrt{F_k F^k}} ~~({\rm triple}),
%%\eeqn
for the optically thin case, and 
\beqn
&&\lambda=-\beta^i+{2 w^2 p^i \pm 
\sqrt{\alpha^2 \gamma^{ii}(2w^2+1)-2(w p^i)^2}
\over 2w^2+1}, \nonumber \\
&&{\rm and}~~\lambda=-\beta^i+p^i~~({\rm double}),
\eeqn
for the optically thick case with $p^i=\alpha V^i/w$ and $l_{(\nu)}=0$. 

\section{Hydrodynamics}

A conservative form of the hydrodynamic equations is derived from
\beqn
&& \nabla_{\alpha} (\rho u^{\alpha})=0,\\
&& \gamma_{\beta i}
\nabla_{\alpha} (T_{\rm fluid}^{\alpha\beta}+T_{\rm rad}^{\alpha\beta})=0, \\ 
&& n_{\beta} \nabla_{\alpha} (T_{\rm fluid}^{\alpha\beta}
+T_{\rm rad}^{\alpha\beta})=0.
\eeqn
The first, second, and third equations are the continuity,
Euler, and energy equations, respectively. For the perfect fluid, 
\beqn
T_{\rm fluid}^{\alpha\beta}=\rho h u^{\alpha}u^{\beta}+P g^{\alpha\beta},
\eeqn
where $h$ is the specific enthalpy defined by $1+\varep+P/\rho$,
and $\varep$ and $P$ are the specific internal energy and pressure. 

For the neutrino-radiation hydrodynamics, it is further necessary to 
solve the continuity equation for leptons or electrons: The 
continuity equation for electrons is written in the form
\beqn
\nabla_{\alpha} (\rho Y_e u^{\alpha})=\rho Q_e,
\eeqn
where $Y_e$ denotes the electron number per nucleon and $Q_e$ the 
electron generation rate, determined by the electron capture and 
neutrino capture by nucleons or nuclei. 

The explicit forms for these equations are
\beqn
&& \pa_t (\sqrt{\gamma} \rho w) + \pa_j (\sqrt{\gamma} \rho w v^j)=0, \\
&&\pa_t (\sqrt{\gamma} j_{i})
+\pa_j[\sqrt{\gamma}(j_{i} v^j + \alpha P \delta^{~j}_{i}) ]
\nonumber \\
&&\hskip 1.2cm 
=\sqrt{\gamma}\Big[-\rho_{\rm h} \pa_i \alpha + j_{k} \pa_i \beta^k
+{\alpha \over 2} S^{jk}\pa_i \gamma_{jk}
- \alpha S^{\alpha} \gamma_{\alpha i} \Big], \label{eq2.1} \\
%%%%%%%%%%%%%%%%%%%%%%%%%%%%%%%%%%%%
&& \pa_t (\sqrt{\gamma} \rho_{\rm h})+\pa_j[\sqrt{\gamma}
\{\rho_{\rm h} v^j + P (v^j+\beta^j)\}] \nonumber \\
&&\hskip 1.2cm 
=\alpha\sqrt{\gamma}[S^{ij} K_{ij}
-\gamma^{ik} j_{i} \pa_k \ln \alpha +S^{\alpha} n_{\alpha}],
\label{eq2.2}\\
&& \pa_t (\sqrt{\gamma} \rho w Y_e) + \pa_j (\sqrt{\gamma} \rho w Y_e
v^j)=\rho Q_e \alpha \sqrt{\gamma}, 
\eeqn
where $v^i=u^i/u^0$, and 
\beqn
&&j_{i}=-T_{\rm fluid}^{\alpha\beta}n_{\alpha}\gamma_{\beta i}=\rho w h u_i,\\
&&\rho_{\rm h}=T_{\rm fluid}^{\alpha\beta}n_{\alpha}n_{\beta}=\rho h w^2 -P, \\
&&S^{jk}=T_{\rm fluid}^{\alpha\beta}\gamma_{\alpha}^{~j}\gamma_{\beta}^{~k}
=\rho h V^j V^k + P\gamma^{jk}. 
%%%&&S^{\alpha}=\int_0^{\infty} d\nu S_{(\nu)}^{~\alpha}.
%%%&&s_0=-n_{\alpha}\nabla_{\beta}T_{\rm rad}^{\alpha\beta},\\ 
%%%&&s_i=-\gamma_{\alpha i}\nabla_{\beta}T_{\rm rad}^{\alpha\beta}. 
\eeqn
As in the equations for the radiation moments $(E_{(\nu)}, F_{(\nu)i})$, 
the Euler and energy equations have conservative forms. We note 
that extension to the magnetohydrodynamic equations is 
straightforward (e.g., Ref.~\citen{SS05}).

To guarantee the conservation of total momentum and energy, it may be 
useful to solve
\beqn
&&\pa_t [\sqrt{\gamma} (j_{i}+F_i)]
+\pa_j[\sqrt{\gamma}\{j_{i} v^j + \alpha (P \delta^{~j}_{i} 
+ P_i^{~j}) -\beta^j F_i \}] \nonumber \\
&&\hskip 0.2cm 
=\sqrt{\gamma}\Big[-(\rho_{\rm h}+E) \pa_i \alpha + (j_{k} + F_k)\pa_i \beta^k
+{\alpha \over 2} (S^{jk} + P^{jk}) \pa_i \gamma_{jk} \Big], \label{eq2.1a} \\
&& \pa_t [\sqrt{\gamma} (\rho_{\rm h}+E)]+\pa_j[\sqrt{\gamma}
\{\rho_{\rm h} v^j + P (v^j+\beta^j) + \alpha F^j-\beta^j E\}] 
\nonumber \\ &&\hskip 1.2cm 
=\alpha\sqrt{\gamma}[(S^{ij} + P^{ij}) K_{ij}
-\gamma^{ik} (j_{i} +F_i) \pa_k \ln \alpha ].
\label{eq2.3a}
\eeqn
We note that $E$, $F_k$, and $P^{ij}$ here denote the 
sum of the contribution from all the neutrino species. 

\section{Slow-motion limit}

Here, we derive the radiation hydrodynamics equations in the case that
(i) the spacetime is flat and (ii) the typical velocity of the matter
field is much smaller than the speed of light. These approximations
are often used in the radiation hydrodynamics with Newtonian gravity.
Here, several additional words are necessary to clarify the condition
(ii). First, we denote the typical time and length scales for the
variation of the matter field by $T$ and $L$, respectively, and the
velocity by $V$. Then, the order of $T$ is equal to $L/V$, and the
acceleration and shear of the matter are of order $V/T \sim V^2/L$.
In the Newtonian approximation for the radiation hydrodynamics, we
take into account all the terms associated with the first order in $V$
relative to the lowest-order term, but neglect the terms more than
second order in $V$; terms of $O(V^2)$ such as acceleration and
$(\pa_i v_j)^2$ are neglected. Because the Newtonian potential is the
quantity of order $V^2$, we also neglect the contribution by this in
the radiation moment equations.

Then, the equations for the radiation moments defined in
the fluid comoving frame, $(J_{(\nu)},~H_{(\nu)}^{~i})$, are
%%%%%%%%%%%%%%%%%%%%
\beqn
&&\pa_t J_{(\nu)} + \pa_i (J_{(\nu)} v^i +H_{(\nu)}^{~i}) - \nu {\pa
(L_{(\nu)}^{~ij} \pa_i v_j) \over \pa \nu} =\kappa_{(\nu)}(J^{\rm
eq}_{(\nu)}-J_{(\nu)}),\label{NewJ}\\ 
&&\pa_t H_{(\nu)}^{~i}+\pa_j (H_{(\nu)}^{~i} v^j + H_{(\nu)}^{~j} v^i
+L_{(\nu)}^{~ij}) - v^i \pa_j H_{(\nu)}^{~j}
-{\pa
\over \pa \nu}(\nu N_{(\nu)}^{~ijk}\pa_j v_k)=-\tilde \kappa_{(\nu)}
H_{(\nu)}^{~i},
\label{NewH}
\eeqn
where we set $u^0=1 + O(V^2)$, $u_0=-1 + O(V^2)$, $u^i=u_i=v^i=v_i$,  
$a^k:=u^{\mu} \nabla_{\mu} u^k=O(V^2)$, and
\beqn
&&N_{(\nu)}^{~ijk}\pa_j v_k= {3\chi -1 \over 2} {J_{(\nu)} H_{(\nu)}^{~i}
H_{(\nu)}^{~j} H_{(\nu)}^{~k} \over (H_{(\nu)}^{~l} H_{(\nu)l})^{3/2}}
\pa_j v_k \nonumber \\
&&~~~~~~~~~~~~~~~~~~~~~~~~
+{3(1-\chi)\over 10}\Big(
H_{(\nu)}^{~i} \pa_j v^j + H_{(\nu)}^{~j} \pa_j v^i + H_{(\nu)}^{~j} \pa_i v_j 
\Big), \\
&&L_{(\nu)}^{~ij}={3\chi -1 \over 2}
{J_{(\nu)} H_{(\nu)}^{~i} H_{(\nu)}^{~j} \over H_{(\nu)}^{~k} H_{(\nu)k}} 
+{1-\chi \over 2} J_{(\nu)} \delta^{ij}.
\eeqn
%%%%%%%%%%%
For simplicity, we here take into account only the neutrino emission,
absorption, and isoenergy scattering.  The derived forms of
Eqs.~(\ref{NewJ}) and (\ref{NewH}) agree with those in the standard
textbooks (e.g.,~Ref.~\citen{Miha}).  On the other hand, the equations
for the radiation moments defined in a laboratory frame,
$(E_{(\nu)},~F_{(\nu)i})$, are
\beqn
&&\pa_t E_{(\nu)} + \pa_i F_{(\nu)}^{~i} - {\pa
(\nu L_{(\nu)}^{~ij} \pa_i v_j) \over \pa \nu} =\kappa_{(\nu)}(J^{\rm
eq}_{(\nu)}-E_{(\nu)} +  F_{(\nu)}^{~i} v_i),\label{LabJ} \\ 
%%%%%%%%%%%%%%%%%%%%%%%
&&\pa_t F_{(\nu)}^{~i}+\pa_j P_{(\nu)}^{~ij}
-{\pa \over \pa \nu}(\nu N_{(\nu)}^{~ijk}\pa_j v_k)\nonumber \\
&& \hskip 3cm 
=-\tilde \kappa_{(\nu)} (F_{(\nu)}^{~i}-P_{(\nu)}^{~ik}v_k)
+[\kappa_{(\nu)} J^{\rm eq}_{(\nu)}+(\tilde \kappa_{(\nu)} - \kappa_{(\nu)}) 
E_{(\nu)}]v^i,\label{LabH}
\eeqn
where
\beqn
&&N_{(\nu)}^{~ijk}\pa_j v_k= {3\chi -1 \over 2} {E_{(\nu)} F_{(\nu)}^{~i}
F_{(\nu)}^{~j} F_{(\nu)}^{~k} \over (F_{(\nu)}^{~l} F_{(\nu)l})^{3/2}}
\pa_j v_k \nonumber \\
&&~~~~~~~~~~~~~~~~~~~~~~~~~~~
+{3(1-\chi)\over 10}\Big(
F_{(\nu)}^{~i} \pa_j v^j + F_{(\nu)}^{~j} \pa_j v^i + F_{(\nu)}^{~j} \pa_i v_j 
\Big), \\
&&L_{(\nu)}^{~ij}={3\chi -1 \over 2}
{E_{(\nu)} F_{(\nu)}^{~i} F_{(\nu)}^{~j} \over F_{(\nu)}^{~k} F_{(\nu)k} } 
+{1-\chi \over 2} E_{(\nu)} \delta^{ij},\\
&&P_{(\nu)}^{~ij}={3\chi -1 \over 2}
{E_{(\nu)} F_{(\nu)}^{~i} F_{(\nu)}^{~j} \over F_{(\nu)}^{~k} F_{(\nu)k} } 
+{3(1-\chi) \over 2}
\Big({E_{(\nu)} \over 3} \delta^{ij}
+F_{(\nu)}^{~i} v^j+F_{(\nu)}^{~j} v^i
-{2\over 3} \delta^{ij} F_{(\nu)}^{~k} v_k
\Big),
\eeqn
and we used
\beqn
&& J_{(\nu)}=E_{(\nu)}-2F_{(\nu)}^{~k} v_k,\\
&& H_{(\nu)}^{~k}=-E_{(\nu)}v^k + F_{(\nu)}^{~k}-P_{(\nu)}^{~kl} v_l.
\eeqn
%%%%%%%%
Again, we note that $\nu$ is the frequency in the {\em fluid rest
frame} (not in the laboratory frame).  As expected, Eqs.~(\ref{LabJ})
and (\ref{LabH}) have a conservative form. 

The hydrodynamic equations are
\beqn
&& \pa_t \rho  + \pa_j (\rho v^j)=0, \\
&&\pa_t (\rho v_i)
+\pa_j (\rho v_i v^j + P \delta^{~j}_{i})
=-\rho \pa_i \phi_{\rm N} 
 \nonumber \\
&& \hskip 1.5cm 
+ \int d\nu \Big(\tilde \kappa_{(\nu)} (F_{(\nu)i}-P_{(\nu)ik}v^k)
-[\kappa_{(\nu)} J^{\rm eq}_{(\nu)}+(\tilde \kappa_{(\nu)} - \kappa_{(\nu)}) 
E_{(\nu)}]v_i\Big), \\
%%%%%%%%%%%%%%%%%%%%%%%%%%%%%%%
&& \pa_t \Big[\rho\Big(\varepsilon + {v^2 \over 2}\Big)\Big]
+\pa_j \Big[\Big(\rho \varepsilon + P + {1\over 2}\rho v^2 \Big)v^j \Big]
\nonumber \\
&& \hskip 0.5cm =-\rho v^i \pa_i \phi_{\rm N} -\int d\nu
\kappa_{(\nu)}(J_{(\nu)}^{\rm eq} - E_{(\nu)} + F_{(\nu)}^{~i} v_i),\\ &&
\pa_t (\rho Y_e) + \pa_j (\rho Y_e v^j)=\rho Q_e,
\eeqn
where $\phi_{\rm N}$ is the Newtonian potential. We note that 
when taking the Newtonian limit of general relativistic 
hydrodynamics equations, the conservative rest-mass density 
$\rho w \sqrt{\gamma}$ is replaced to $\rho$. 
The total energy equation is written as
\beqn
&& \pa_t \Big[\rho\Big(\varepsilon + {v^2 \over 2}\Big)+ E\Big]
+\pa_j \Big[\Big(\rho \varepsilon + P + {1\over 2}\rho v^2\Big)v^j +F^j \Big]
=-\rho v^i \pa_i \phi_{\rm N} . 
\eeqn

\section{Summary}

We derived a truncated moment formalism for general relativistic
radiation hydrodynamics modifying the Thorne's original
formalism~\cite{Kip}.  The equations for the radiation field are
written for the variables defined in the laboratory frame as well as
in the fluid local rest frame, although the argument angular frequency
for the radiation moments is always the frequency measured in the
fluid local rest frame.  In the former case, the equations are written
in a conservative form (for $E_{(\nu)}$ and $F_{(\nu)i}$) and
essentially the same as those for the hydrodynamic equations in
general relativity. Thus, they seem to be useful for a well-resolved
numerical simulation.

The source terms are written, focusing on the neutrino transfer in the
assumption that anisotropy of the scattering kernel is small.  Then, a
formalism for the radiation hydrodynamics in numerical relativity is
derived in a closed form, assuming a physically reasonable closure
relation among the radiation stress tensor, energy density, and energy
flux. As long as the radiation field is not extremely anisotropic in
the fluid rest frame (in the optically thick medium), the employed
approximation should work well. One merit in the present formalism is
that we do not have to perform any coordinate transformation when
computing the source term, because the angular frequency for the
radiation field is defined in the fluid local rest frame. 

The derived equations constitute wave equations for the radiation
field. The closure relation and variable Eddington factor are
appropriately chosen so that the characteristic speed is smaller than
the speed of light in the free-streaming and grey regions. We also
notice that (i) for the derivation of the basic equations for the
radiation field, we do not assume that the fluid velocity is much
smaller than the speed of light, and (ii) with the chosen closure
relation, the effect associated with the fluid motion (fluid
expansion, acceleration, and shear) may be taken into account. Thus,
the derived formalism can be employed for the radiation field
associated with a fast motion, e.g., a fluid moving in the vicinity of
a black hole.

In this formalism, we need to solve 3+1+1 equations (3 is space, 1 is
time, and frequency space) of $(E_{(\nu)}, F_{(\nu)i})$ or 3+1
equations of $(E, F_i)$ for the radiation part. For both cases, the
equations for these variables are written in a conservative form, and
hence, conservation of mass and moments is likely to be well 
achieved in the formalism with these quantities.  For the 3+1+1 case,
the equations, including absorption, emission, and scattering term for
neutrinos, are written in the closed form, and hence, we do not have
to assume anything further. For the 3+1 case, computational costs will
be saved significantly, but we have to impose several additional
conditions when performing the frequency integral for the source term:
We have to assume certain functions for $J_{(\nu)}$ and
$H_{(\nu)}^{~\alpha}$ (e.g.,~Ref.~\citen{Gon}), for which a physically
appropriate assumption is required.

The truncated moment formalism may be a starting point for upgrading
the current leakage scheme for general relativistic radiation
hydrodynamics (e.g., Ref.~\citen{SS} for a review). The leakage scheme
is often used for phenomenologically incorporating radiation cooling
and for a relatively inexpensive radiation hydrodynamic simulation.
The method is usually quite phenomenological: One first determines
optically thick and thin regions, respectively, using a rather
approximate prescription. Then, for the optically thick region, one assumes
that the radiation escapes in a diffusive manner and for the optically
thin region, the radiation escapes freely. In general relativistic
leakage schemes \cite{SS}, one incorporates the cooling effect in the
right-hand side of the hydrodynamics equations
($\nabla_{\alpha}T^{\alpha\beta}=-S^{\beta}$), and in addition, an
equation for radiation four-vector field is evolved for the optically
thin region. Namely, the basic equations are quite similar to those
derived in this paper. What is different is on the treatment for the
source term $S_{(\nu)}^{~A_k}$; in the leakage scheme proposed so far,
the source term is determined in a quite phenomenological manner.  If
$S_{(\nu)}^{~A_k}$ is approximated in a more strict manner starting from
the basic equations derived in this paper, it will be possible to
derive a better and well-funded leakage scheme. Furthermore, it will
be possible to incorporate the frequency-dependent effect as well as
neutrino heating.  Such work is left for the future.

Finally, the truncated moment formalism derived here may be used for
the transfer of photons by exchanging the source terms (if we may
assume that anisotropy of the scattering kernel is small).  For
example, this formalism will be useful for studying an accretion flow
in the vicinity of stellar-mass and supermassive black holes in
general relativity.

\vskip 3mm
\begin{center}
{\bf Acknowledgments}
\end{center}
\vskip 3mm

We thank T. Muranushi for valuable discussion and comments.  This work
was supported by Grant-in-Aid for Scientific Research (21340051), 
by Grant-in-Aid for Scientific Research on Innovative Area (20105004) 
of Japanese MEXT, by JSPS research fellowship, and 
by Grant-in-Aid for Young Scientists (B) 22740178.

\appendix

\section{Stationary radiation in the spherical dilute medium}

In this appendix, we show the solution of a radiation field in the
Bondi flow \cite{ST} composed of a dilute medium (i.e., optically thin
approximation is assumed to work) and illustrate that gravitational
redshift and Doppler effects are appropriately taken into account in
the radiation spectrum observed at infinity in our formalism.

As discussed in \S~\ref{sec:trun}, in the optically thin region 
(where $S_{(\nu)}^{A_k}=0$), the radiation moments may be written by
\beqn
H_{(\nu)}^{~\alpha}=J_{(\nu)} \ell^{\alpha},~~~~
L_{(\nu)}^{~\alpha\beta}=J_{(\nu)} \ell^{\alpha}\ell^{\beta},\label{eqA1}
\eeqn
where $\ell^{\alpha}$ denotes a unit spatial vector for which the
spatial component is composed only of the radial component.
Substitution of these relations into Eq.~(\ref{eqQ0}) with
$S_{(\nu)}^{~\alpha}=0$ yields 
\beqn
\nabla_{\alpha}(J_{(\nu)} l^{\alpha})-\nu {\pa J_{(\nu)} \over \pa \nu}
\ell^{\alpha}l^{\beta} \nabla_{\beta} u_{\alpha}=0,\label{eqap}
\eeqn
where $l^{\alpha}:= u^{\alpha} + \ell^{\alpha}$ is a null
vector. Substitution, in addition, of
$N_{(\nu)}^{~\alpha\beta\gamma}
=J_{(\nu)}\ell^{\alpha}\ell^{\beta}\ell^{\gamma}$
into Eq.~(\ref{eqQ1}) yields the same equation as Eq.~(\ref{eqap}).

For a solution of the four velocity, $u^{\alpha}$, we here consider a
stationary spherical accretion flow (Bondi accretion flow) in the
spacetime of a spherical black hole of mass $M$. We choose the line
element in the Kerr-Schild coordinates,
\beqn
ds^2=-\Big(1 - {2GM \over r}\Big)d{\bar t}^2 
+{4M \over r} d{\bar t}dr + \Big(1 + {2GM \over
r}\Big)dr^2+r^2 (d\theta^2 + \sin^2\theta d\varphi^2), 
\label{KS}
\eeqn
in which the coordinate singularity at $r=2GM$ does not give 
any messy problem. We note however that an analytic solution 
is also easily derived in the Schwarzschild coordinates. 

We denote the infall velocity by $u^r=-u(r) < 0$ (cf. for the wind
solution $u^r > 0$), and note the relations $\ell_{t}=u$ and
$\ell_r=u^t$, where $-u_t=u^t(1-2GM/r)+2G M u/r=\sqrt{1+u^2-2GM/r}$.
Thus, $l^r=-u + \sqrt{1+u^2-2GM/r} > 0$, and $l^t=l^r(r+2GM)/(r-2GM)$.
Using these relations, we finally reach 
\beqn
\ell^{\alpha}l^{\beta} \nabla_{\beta} u_{\alpha}={d l^r \over dr}.
\eeqn
Substitution of this relation into Eq.~(\ref{eqap}) gives 
\beqn
l^r {\pa J_{(\nu)} \over \pa r}+{J_{(\nu)} \over r^2}{d (r^2 l^r) \over dr}
=\nu{\pa J_{(\nu)} \over \pa \nu} {d l^r \over dr},
\eeqn
and assuming that $J_{(\nu)} >0$ and $l^r >0$, we obtain 
\beqn
{\pa y_{(\nu)} \over \pa r}=\nu {\pa y_{(\nu)} \over \pa \nu} 
{d \ln l^r \over dr},\label{eqyy}
\eeqn
where $y_{(\nu)}:=J_{(\nu)} l^r r^2$. For the case that 
$l^r$ is a monotonically increasing function of $r$ 
(this is the case for the typical problem), 
Eq.~(\ref{eqyy}) is rewritten to give
\beqn
{\pa y_{(\nu)} \over \pa \ln l^r}={\pa y_{(\nu)} \over \pa \ln \nu}.
\eeqn
Thus, $y_{(\nu)}(r)$ constitutes a wave equation for the arguments 
$(\ln l^r, \ln \nu)$, and therefore, the general solution can be derived as
\beqn
y_{(\nu)}=F(l^r \nu),~~{\rm or}~~J_{(\nu)}={F(l^r \nu) \over l^r r^2}, 
\label{solution}
\eeqn
where $F(x)$ is an arbitrary function of $x$. Here, $l^r \rightarrow
1$ for $r \rightarrow \infty$. Namely, at infinity, the observed
spectrum is
\beqn
J_{(\nu)} = {F(\nu) \over r^2}. 
\eeqn
On the other hand, for the finite value of $r$, $l^r$ is smaller than
unity.  This implies that the radiation spectrum should be
homogeneously (irrespectively of the value of $\nu$) shifted to the
lower frequency side during outgoing propagation. The redshift factor
is given by $l^r$; an observed radiation with frequency $\nu$ at
infinity is originally emitted at a finite radius, $r_{\rm emit}$,
with frequency $\nu_{\rm emit}=\nu/l^r(r_{\rm emit}) > \nu$.

Equation~(\ref{solution}) indeed captures gravitational redshift and
Doppler effects. This is clearly found by taking the slow-motion and
weak-gravitation approximation for $l^r$ as
\beqn
l^r \approx 1 - u +{u^2 \over 2}-{GM \over r}. 
\eeqn
The first, second, and third terms denote the Doppler, second-order 
Doppler, and gravitational redshift effects, respectively. 

%$E_{(\nu)}$ is calculated to give
%\beqn
%E_{(\nu)}=J_{(\nu)}\Big(1+{2GM \over r}\Big)^{-1}(l^t)^2
%\eeqn

Integration of $J_{(\nu)}$ by $\nu$ gives
\beq
J=\int d\nu J_{(\nu)}={L_0 \over (l^r r)^2},
\eeq
where $L_0$ denotes a total flux
\beqn
L_0=\int d\nu F(\nu).
\eeqn
Thus $l^r r$ may be regarded as a luminosity distance.  Note that for
$r \rightarrow 2GM$, $l^r \rightarrow 0$.  Thus, $J \rightarrow
\infty$ at $r=2GM$; the solution is similar to a solution in the flat
spacetime in which a ``point'' source exists at origin (here which is
located at $l^r r =0$).

Finally, we point out that the solution derived here will be used as a
test-bed problem for checking the reliability of a radiation
hydrodynamic code based on the truncated moment formalism 
(note that with Eq.~(\ref{eqA1}) the closure relation holds).  We also
note that the solution given here holds not only for the Bondi flow
but also any solution in the stationary, spherically symmetric spacetime
as long as $l^r$ is a monotonic function of $r$.

\section{Radiation flow in the spherical dilute medium}

Next, we analyze a time-dependent spherically symmetric radiation flow
in the Schwarzschild spacetime. Again, we assume that neutrinos
propagate in the optically thin medium.  The purpose of this section is
to clarify a nature of the closure relations (\ref{clos1}) and
(\ref{clos2}). For this, we ignore the frequency-dependent effects and
analyze Eqs.~(\ref{eq1.3a}) and (\ref{eq1.3b}). For the background
metric, we again adopt Eq.~(\ref{KS}). In this case, the necessary
geometric quantities are
\beqn
&&\alpha=\Big(1+{2M \over r}\Big)^{-1/2},~~
\beta^r={2M \over r+2M},~~
\gamma_{rr}=1+{2M \over r},\nonumber \\
&&{\rm and}~~
K_{rr}=-{2M(r+M) \over r^{5/2} (r+2M)^{1/2}}, 
\eeqn
where we use the units of $G=1$ (or we may say that $GM$ 
is replaced to $M$). 
Because of the spherical symmetry, we only need to
consider the radial component of radiation moments, $F^r$.  For the
following, we define $F := F^r \gamma_{rr}^{1/2}$.  Then the
condition $g_{\mu\nu}T_{\rm rad}^{\mu\nu}=0$ for the closure 
relations (\ref{clos1}) and (\ref{clos2}) is written as 
\beq
E=|F| \label{cond99}
\eeq

For the closure relation (\ref{clos1}), the equations for $E$ and $F$ 
are
\beqn
&& \dot e -{2M \over r+2M}e'+ {r \over r+2M} f'
+{2M(2r+M) \over r(r+2M)^2}e +{3M \over (r+2M)^2} f=0,\\
&& \dot f - {2M \over r+2M} f'+{r \over r+2M}e'
+{2M(2r+M) \over r(r+2M)^2}f +{3M \over (r+2M)^2} e=0,
\eeqn
where $e=E r^2 \gamma_{rr}^{1/2}$ and $f=F r^2\gamma_{rr}^{1/2}$. 
The dot ($\dot e$) and dash ($e'$) denote $\pa_t e$ and $\pa_r e$, 
respectively. Defining $u_{\pm}=e \pm f$, we obtain two
independent equations
\beqn
&&\dot u_+ + {r-2M \over r+2M} u_+' + {M(7r+2M) \over r(r+2M)^2}
u_+=0, \label{B5} \\
&&\dot u_- - u_-' + {M \over r(r+2M)} u_-=0, \label{B6}
\eeqn
and $e=(u_+ + u_-)/2$ and $f=(u_+ - u_-)/2$. This implies that in the
absence of $u_+$ or $u_-$, the condition (\ref{cond99}) is satisfied,
but in general, it is not. In particular, for the point which
satisfies $f=0~(u_+=u_-)$, one of the characteristic speed becomes
infinity (see Eq.~(\ref{eq:chara1})). \footnote{In the analysis of
spherically symmetric flow here, the characteristic speeds are
$(r-2M)/(r+2M)$ and $-1$. However, if we solve the equation in the
Cartesian or cylindrical coordinates, the extra characteristic speed
(\ref{eq:chara1}) appears.} Thus this closure relation should be
prohibited for such a situation (this is resolved in an appropriate
choice of the variable Eddington factor shown in \S 6.3).

For the closure relation (\ref{clos2}), the equations for $e$ and $f$ 
are
\beqn
&& \dot e -{2M \over r+2M}e'+ {r \over r+2M} f'
+{2M \over (r+2M)^2}e +{3Mr+2M(r+M)s \over r(r+2M)^2} f=0,\\
&& \dot f - {2M -r s \over r+2M} f'
+{2M(M+2r+rs) \over r(r+2M)^2}f +{M \over (r+2M)^2} e=0,
\eeqn
where $s=1~(-1)$ for $F^r >0~(<0)$. 
Defining $u=e-s f=(E-s F)r^2\gamma_{rr}^{1/2}$, we obtain
\beqn
&&\dot u - {2M \over r+2M} u' + {2M-s M \over (r+2M)^2} u=0,\\
&& \dot f - {2M -r s \over r+2M} f'
+{M(2M+4r+3rs) \over r(r+2M)^2}f +{M \over (r+2M)^2} u=0.\label{B10}
\eeqn
Thus, there are also two components: One is determined by $u$ which is a
mode of physically zero characteristic speed because the coefficient
of the transport term is equal to $-\beta^r$.  The other is associated
with $f$, which is an outgoing or ingoing mode and obeys the same
equation as that of $u_+$ and $u_-$ for $f >0 $ and $f < 0$,
respectively. $u$ is regarded as an unphysical mode because it is the
measure of deviation from the condition (\ref{cond99}); if $u=0$ is
satisfied, we can follow only the physical mode, but this will not be
in general the case, in particular in the near zone.  The important
fact, however, is that $u$ does not propagate outward and damps
exponentially with time in the absence of the source term.  This
implies that in the zone distant from the source, $u$ will be zero
because the emission source should be zero there. Thus, in the distant
optically thin zone, the condition (\ref{cond99}) is likely to be
satisfied. 

It will be useful to give the solutions of Eqs.~(\ref{B5}) and
(\ref{B6}): The general solutions for these are written as
\beqn
&&u_+=\Big[{r(r+2M)^3 \over (r-2M)^4}\Big]^{1/2}g_+(t-r_*),\\
&&u_-=\Big[{r \over r+2M}\Big]^{1/2}g_-(t-r),
\eeqn
where $g_{\pm}$ are arbitrarily functions and $r_*$ is a retarded time
\beqn
r_*=\int dr {r+2M \over r-2M}=r + 4M \ln \Big({r-2M \over M}\Big).
\eeqn

\section{Numerical experiment for free evolution}

To confirm that the closure relation and variable Eddington factor
(\ref{chi1}) described in \S 6 work well in the optically thin medium, 
we numerically solve radiation field equations (\ref{eq1.3a}) and 
(\ref{eq1.3b}) on a Bondi flow of a Schwarzschild spacetime. 
The closure relation is written as
%%%%%%%%%%%%%%%%%%%%
\beqn
P^{ij}={3\chi - 1 \over 2}E {F^i F^j \over \gamma_{kl}F^k F^l}
+{3(1-\chi) \over 2}\biggl(J {\gamma^{ij}+4V^i V^j \over 3}
+H^i V^j + H^j V^i \biggr),
\eeqn
%%%%%%%%%%%%%%%%%%%%
where $\chi$ is assumed to be a function of ${\bar
F}=|F|/E=\sqrt{\gamma_{kl}F^k F^l}/E$.  The source terms are set to be
zero ($S^{\alpha}=0$) for simplicity.  Numerical simulation was
performed assuming the axial and equatorial plane symmetries. As in
Appendix A and B, the Kerr-Schild coordinates are adopted, and the
same Bondi solution as in Ref.~\citen{SS05} is employed.  The basic
equations are essentially the same as those solved in general
relativistic hydrodynamic simulation.  We employ the same scheme as
used in Ref.~\citen{SS05} for a solution of $E$ and $F_{k}$.
Specifically, the transport term is handled using a Kurganov-Tadmor
scheme \cite{KT} with a piecewise parabolic reconstruction for the
quantities of cell interfaces.  The fourth-order Runge-Kutta method is
employed for the time integration. The characteristic speed is not
analytically computed for the general form of $P_{ij}$. Thus, we
simply write it in the linear combination form
%%%%%%%%%%%%%%%%%%%%
\beqn
\lambda={3\chi - 1 \over 2}\lambda_{\rm thin}
+{3(1-\chi) \over 2}\lambda_{\rm thick}.
\eeqn
%%%%%%%%%%%%%%%%%%%%
For the case that the relation, $E < |F|$, is accidentally realized at
a point, we set $\chi=1$, and $\lambda_{\rm thin}$ is limited to be
smaller than unity.

With the time evolution, the radiation fields flow away from the
computational domain. To handle this correctly, an outgoing boundary
condition is imposed for the outer boundaries, and inside the radius
$r \leq 1.8M$, we artificially set $E=F_{k}=0$.

First, we consider the solution for $E=|F|$ derived in Appendix B. 
In this case, ${\bar F}$ is always unity, and thus, $\chi=1$ 
always holds.  For the outgoing flow $F=E$, the solution is written as
%%%%%%%%%%%%%%%%%%%%
\beqn
E \gamma_{rr}^{1/2}=F \gamma_{rr}^{1/2}
={1 \over 2}\biggl[{(r+2M)^3 \over r^3(r-2M)^4}\biggr]^{1/2}g_+(t-r_*),
\eeqn
%%%%%%%%%%%%%%%%%%%%
and for the ingoing flow $F=-E$, 
%%%%%%%%%%%%%%%%%%%%
\beqn
E \gamma_{rr}^{1/2}=-F \gamma_{rr}^{1/2}={1 \over 2}[r^3(r+2M)]^{-1/2}g_-(t-r).
\eeqn
%%%%%%%%%%%%%%%%%%%%
We choose a form of a wave packet as
%%%%%%%%%%%%%%%%%%%%
\beqn
g_{+}(r_*)=\exp[-(r_*-r_{*0})^2/8M^2],~~~g_{-}(r)=\exp[-(r-r_{0})^2/8M^2].
\eeqn
%%%%%%%%%%%%%%%%%%%%
Numerical simulations were performed for $r_{*0}=r_*(r=6M)$ and
$r_0=32M$. The computational domain covers a region $[0:40M]$ both for
$x$ and $z$ with a uniform grid spacing $0.1M$. 

Figure 1 plots the evolution of $E
\gamma_{rr}^{1/2}(r-2M)^2(1+2M/r)^{-3/2}$ for the outgoing solution
(left) and $E \gamma_{rr}^{1/2} r^{3/2}(r+2M)^{1/2}$ for the ingoing
solution (right).  This shows that besides a small phase error, the
numerical solutions reproduce the exact solutions.  

%%We find that the propagation of the wave packet is always slightly
%%delayed irrespective of the grid resolution.  This seems to be due to
%%our choice of the variable Eddington factor method.

\begin{figure}[t]
%\vspace{-4mm}
\begin{center}
\epsfxsize=2.4in
\leavevmode
\epsffile{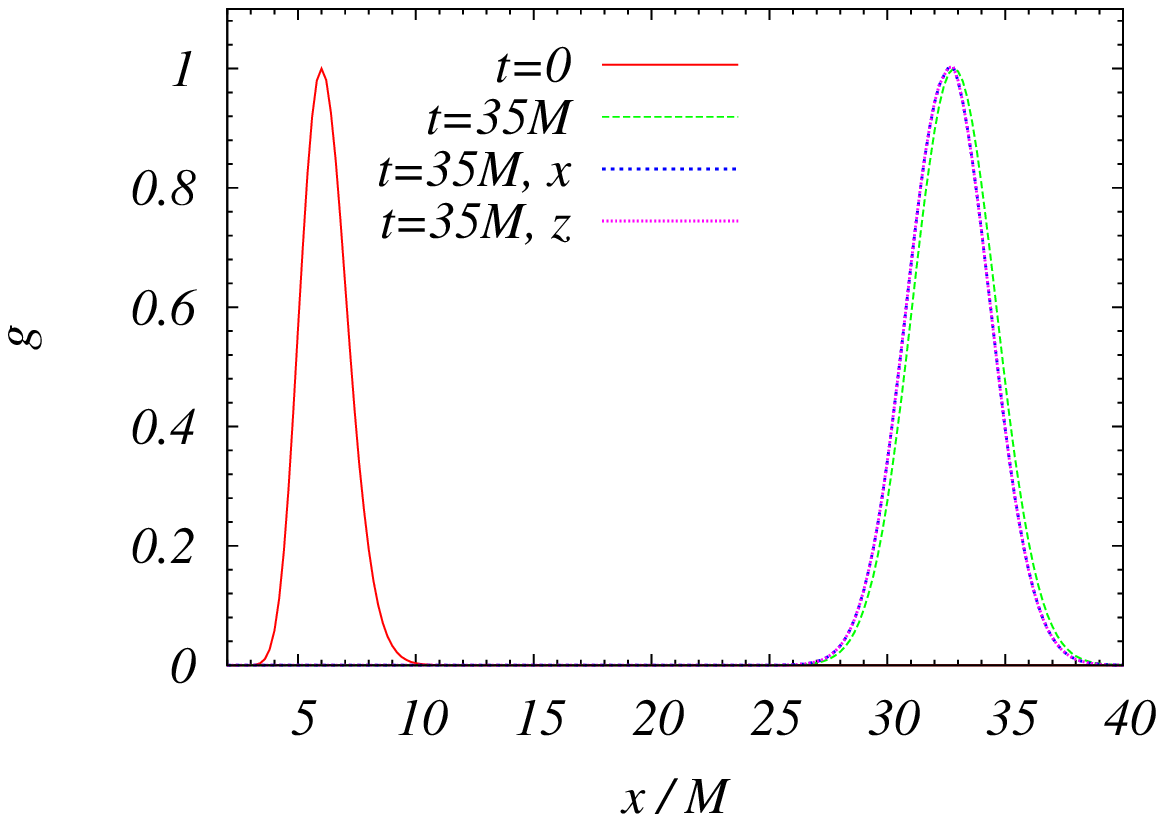}
\epsfxsize=2.4in
\leavevmode
~~~~\epsffile{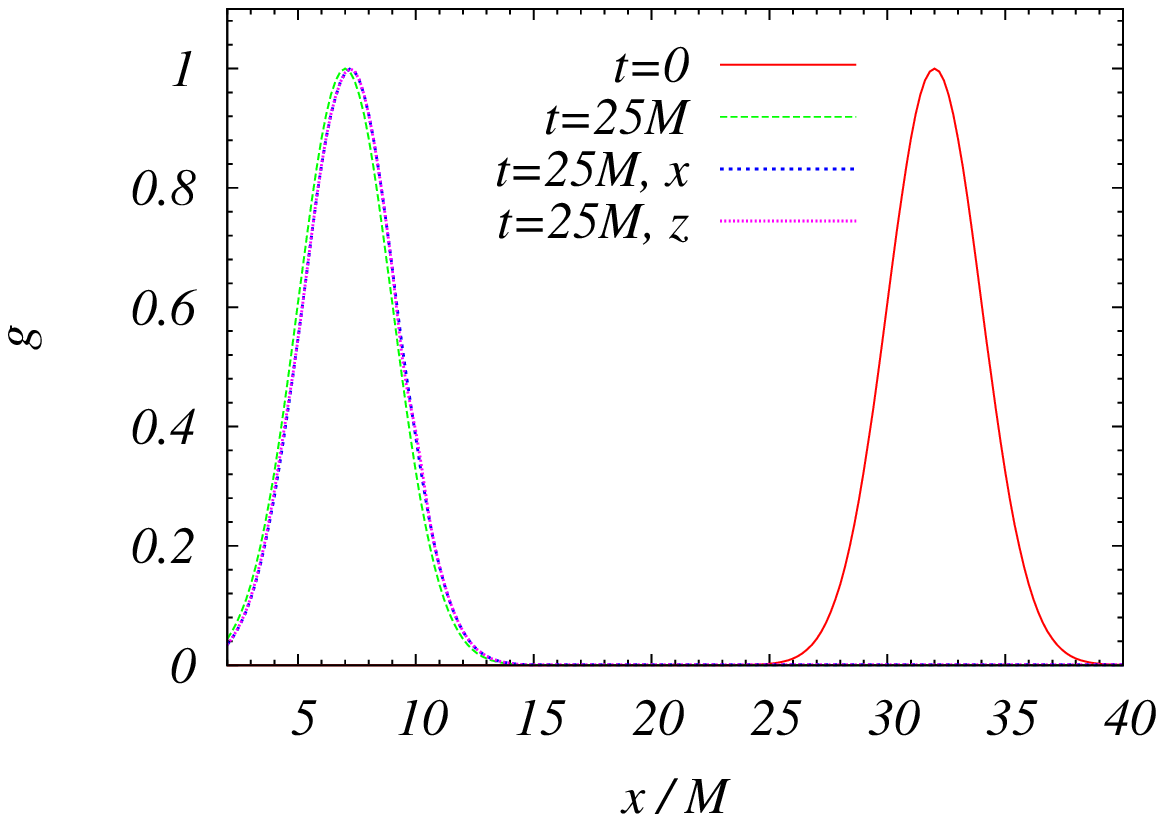}
\caption{Evolution of outgoing (left) and ingoing (right) solutions. 
For the outgoing solution, the profiles of 
$E \gamma_{rr}^{1/2}(r-2M)^2(1+2M/r)^{-3/2}$ at $t=0$ and $t=35M$ 
along the $x$ and $z$ axes are plotted. 
For the ingoing solution, the profiles of 
$E \gamma_{rr}^{1/2} r^{3/2}(r+2M)^{1/2}$ at $t=0$ and $t=25M$ 
along the $x$ and $z$ axes are plotted. 
The solid and dashed curves denote the numerical and 
exact solutions. The numerical solutions for $x$ and $z$ axes 
agree approximately. 
} 
\label{fig1}
\end{center}
\end{figure}

We also performed a simulation for $F=0$ at $t=0$ with 
\beqn
E \gamma_{rr}^{1/2}=\exp[-(r-10M)^2/8M^2].\label{mn2}
\eeqn
In this case, $\chi=1/3$ at $t=0$, but with the time evolution, 
$|F|$ becomes nonzero and $\chi$ becomes larger than $1/3.$ 
Because the variable Eddington factor $\chi$ is varied with 
the evolution, we do not have the exact solution. 
The purpose is to test if our formalism allows a 
stable numerical solution. 

Figure 2(a) plots the time evolution of the wave packet.  Here we plot
$E\gamma_{rr}^{1/2}r^2$ along $x$ and $z$ axes (two results
approximately agree and cannot be distinguished in the figure).  After
the evolution starts, the wave packet is split into outgoing and
ingoing parts. Both modes propagate smoothly with no trouble. Figure
2(b) plots the evolution of $F/E$.  This is initially zero, but with
the free propagation, it approaches to unity: At $t/M=5$, 20, and 50,
$F/E$ is larger than 0.8 for $15 \alt x/M \alt 25$, $15 \alt x/M \alt
/35$, and $x/M \agt 5$, respectively.  No problem is found for the
propagation, and thus, as far as the numerical issues are concerned,
the closure relation and variable Eddington factor employed here have
no problem.

\begin{figure}[th]
%\vspace{-4mm}
\begin{center}
\epsfxsize=2.4in
\leavevmode
(a)\epsffile{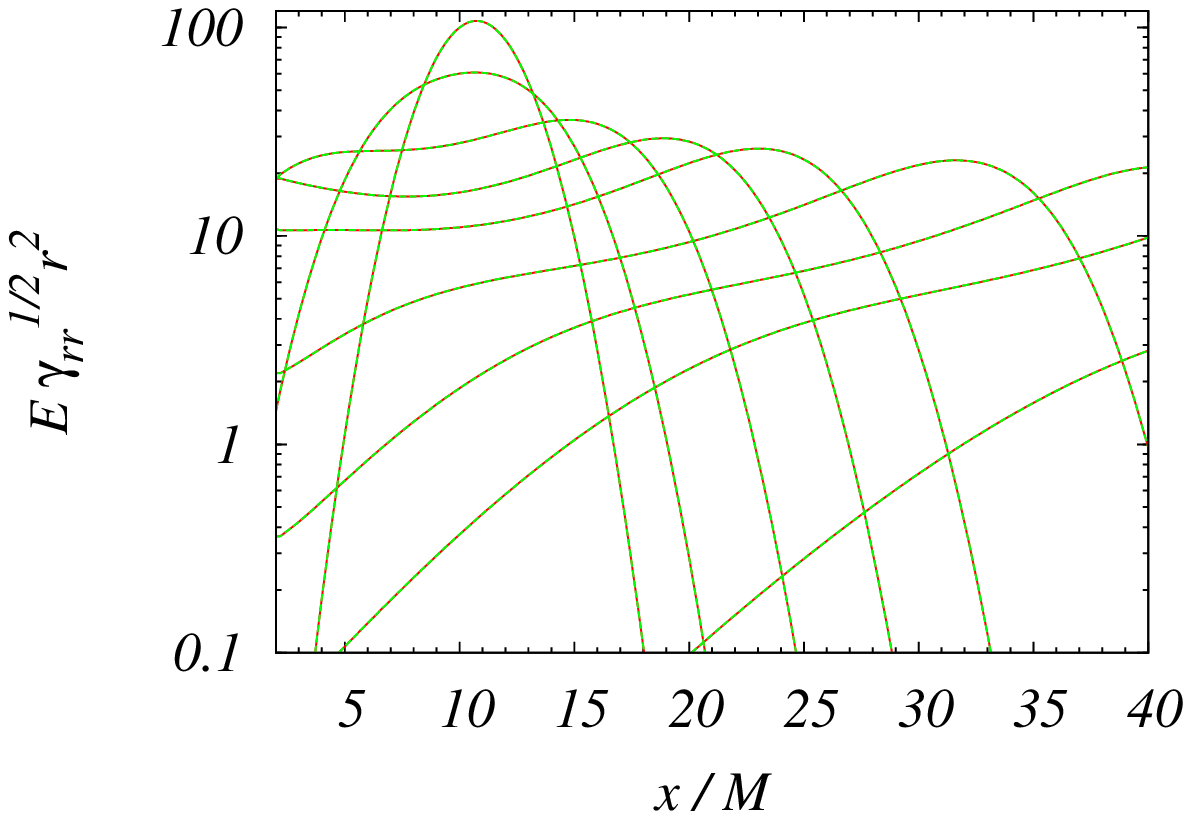}
\epsfxsize=2.4in
\leavevmode
~~~~(b)\epsffile{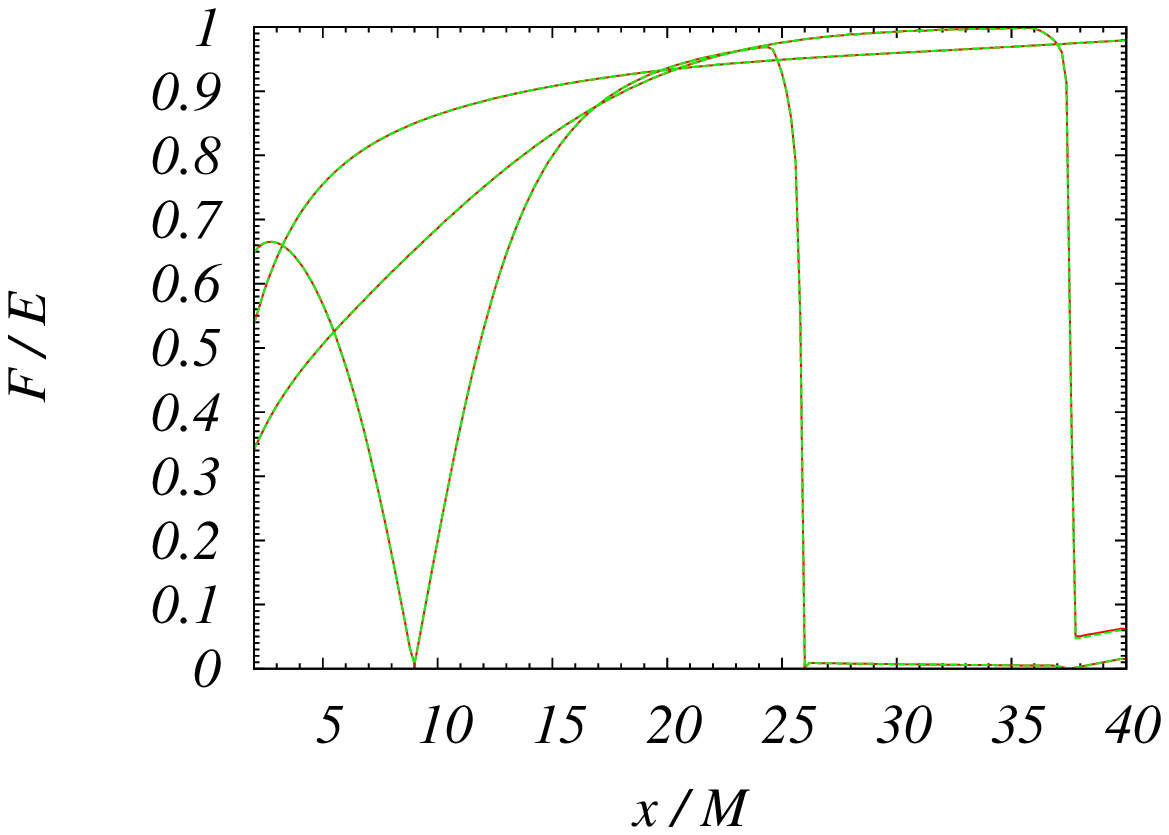}
\caption{(a) Evolution of the wave packet along the $x$ 
and $z$ axes for the initial condition (\ref{mn2}) with $F=0$. 
The profiles are shown for $t/M=0$, 5, 10, 15, 20, 30, 50, and 70. 
(b) $F/E$ along the $x$ and $z$ axes for $t/M=5$, 20, and 50. 
For both figures, the results for $x$ and $z$ axes agree
approximately. 
} 
\label{fig2}
\end{center}
\end{figure}

%%%%%%%%%%%%%%%%%%%%%%%%%%%%%%%%%%%%%%%%%%%%%%%%%%%%%%%%%%%%%

\end{document}